\documentclass[10pt]{article}
\usepackage[hmargin=1.25in,vmargin=1.25in]{geometry}
\usepackage{setspace} 
\usepackage{amsfonts,amsmath,amssymb,amsthm}
\usepackage{xcolor}
\usepackage[colorlinks,linkcolor=teal,citecolor=magenta]{hyperref}
\usepackage{natbib} 
\usepackage{etoolbox}
\usepackage{enumerate}
\usepackage{booktabs}

\allowdisplaybreaks
\renewcommand{\qed}{\rule{2mm}{2mm}}
\newcommand{\indep}{\perp \!\!\! \perp}

\usepackage{accents}
\newcommand{\ubar}[1]{\underaccent{\bar}{#1}}

\DeclareMathOperator*{\argmax}{argmax}

\newtheorem{theorem}{Theorem}[section]
\newtheorem{lemma}{Lemma}[section]
\newtheorem{corollary}{Corollary}[section]
\theoremstyle{definition}
\newtheorem{example}{Example}[section]
\newtheorem{remark}{Remark}[section]
\newtheorem{assumption}{Assumption}[section]

\AtEndEnvironment{remark}{~\qed}
\AtEndEnvironment{example}{~\qed}

\newcommand{\TABLESPACING}{1.2}
\newcommand{\spacingset}[1]{\setstretch{#1}}
\hypersetup{pdftitle={Sharp Testable Implications of Encouragement Designs},
  pdfauthor={Yuehao Bai, Shunzhuang Huang, Max Tabord-Meehan}}

\begin{document}
\author{
Yuehao Bai \\
Department of Economics\\
University of Southern California \\
\url{yuehao.bai@usc.edu}
\and
Shunzhuang Huang \\
Booth School of Business\\
University of Chicago\\
\url{shunzhuang.huang@chicagobooth.edu}
\and
Max Tabord-Meehan\\
Department of Economics\\
University of Toronto \\
\url{m.tabordmeehan@utoronto.ca}
}

\title{Sharp Testable Implications of Encouragement Designs\thanks{The first draft of this paper was dated November 14, 2024. We thank Tim Armstrong, Stephane Bonhomme, Deniz Dutz, Désiré Kédagni, Toru Kitagawa, Simon Lee, Lihua Lei, Matt Masten, Ismael Mourifié, Kirill Ponomarev, Azeem Shaikh, Alex Torgovitsky, Atom Vayalinkal, Edward Vytlacil, Yuanyuan Wan, as well as many workshop participants for helpful comments.}}

\date{October 26, 2026}
\hypersetup{pageanchor=false}
\maketitle
\vspace{-2em}

\begin{abstract}
This paper studies a potential outcome model with a general outcome, a discrete multi-valued treatment, and a discrete multi-valued instrument. We derive sharp, closed-form testable implications for a class of restrictions on potential treatments where each value of the instrument encourages at most one unique treatment choice; such restrictions serve as the key identifying assumption in several prominent recent empirical papers. Borrowing the terminology used in randomized experiments, we call such a setting an encouragement design. The testable implications are inequalities in terms of the conditional distributions of choices and the outcome given the instrument. Through a novel constructive argument, we show these inequalities are sharp in the sense that any distribution of the observed data that satisfies these inequalities is compatible with this class of restrictions on potential treatments. Based on these inequalities, we propose tests of the restrictions. In an empirical application, we show some of these restrictions are violated and pinpoint the substitution pattern that leads to the violation.
\end{abstract}



\thispagestyle{empty}
\newpage
\setcounter{page}{1}
\hypersetup{pageanchor=true}

\section{Introduction}
The analysis of potential outcome models with a discrete, multi-valued treatment and a discrete, multi-valued instrument has gained considerable interest recently. In the setting where both the treatment and the instrument are binary, the well-known monotonicity assumption of \cite{imbens1994identification} plays a key role in the identification of causal effects. Beyond this setting, the most natural extension of the monotonicity assumption for modeling choice behavior is arguably to assume that each value of the instrument (e.g., subsidy or voucher towards a program) increases the appeal of at most one unique treatment choice (e.g., the corresponding program). In fact, such an assumption on choice behavior serves as the key identifying assumption in several prominent recent empirical papers \citep[see, for instance,][]{kirkeboen2016field,kline2016evaluating}. Borrowing the terminology used in randomized experiments, we call such a setting an encouragement design \citep[][]{powers1984effects,holland1988causal}.

Under instrument exogeneity and an exclusion restriction, we derive sharp, closed-form testable implications, in the form of inequalities on the conditional distributions of the outcome and the treatment given the instrument, that characterize when the distribution of the observed data is consistent with an encouragement design. The outcome in our result could be vector-valued with a finite or continuous support. We show through a novel constructive argument that these inequalities are sharp in the sense that they exhaust all the information in the model. Because the inequalities are in closed form, we can test them using existing methods designed for moment inequalities, including but not limited to methods designed for projection inference in linear systems. Moreover, by applying multiple hypothesis testing methods \citep[for instance,][]{romano2018multiple} to these closed-form inequalities, we can formally relate violations of the model to specific choice patterns. Finally, we show that the latent distribution constructed to demonstrate sharpness simultaneously attains the sharp lower bounds of a wide range of parameters, including the probabilities of a large class of substitution patterns. In an empirical application to \cite{behaghel2013robustness,behaghel2014private}, we reject an assumption imposed in the original papers as a harmless ``normalization'' and detect a violated inequality associated with a specific substitution pattern. Using our result on sharp lower bounds, we further quantify the minimum fraction of individuals exhibiting this substitution pattern. In simulations, we find that testing our sharp inequalities can lead to higher power and substantially lower computational cost than tests based on a latent formulation of the problem as a linear program, especially when the support of the outcome is not too small.


To motivate our assumptions on potential treatments, suppose each person receives a voucher (the instrument) towards one of three preschool programs (the treatment choice). We assume that the voucher only increases the appeal of the corresponding program, so receiving the voucher should not change the comparison among the remaining programs. Therefore, for each person, there exists a ``default'' choice (which may differ across people), which is the choice they would have made if the instrument didn't exist; when the instrument equals $j$, then they choose either treatment $j$ or the default choice. These restrictions on potential treatments lead to a set of intuitive inequalities on conditional distributions given the instrument. To the best of our knowledge, these inequalities are new to the literature beyond the setting where both the treatment and the instrument are binary. 

To accommodate a larger class of empirical examples, we allow some choices to have no encouragement. Separately, we also allow a ``base state'' of the instrument which does not encourage any choice. Examples of settings in which both features are needed include \cite{kline2016evaluating} and \cite{kirkeboen2016field}.
In \cite{kline2016evaluating}, the treatment takes three values: no preschool, other preschools, Head Start; and the instrument is a binary indicator of whether the household receives an offer for Head Start. In this case, not receiving the Head Start offer is assumed to be a base state which does not change the appeal of any choice, and the values of no preschool and other preschools do not have corresponding encouragements. Although assuming the existence of a base state is reasonable in \cite{kline2016evaluating}, it is otherwise often motivated as a harmless ``normalization'' that the value of the control status is not affected by the instrument. As we demonstrate in the empirical application to \cite{behaghel2013robustness,behaghel2014private}, however, this ``normalization'' is not innocuous and is potentially refutable in the data. Finally, we also extend our results to settings with one-sided noncompliance, which is often assumed when the instrument represents eligibility to a program; see \cite{kirkeboen2016field} and \cite{angrist2009incentives} for examples.

For the setting with a binary treatment, a binary instrument, and a binary outcome, \cite{balke1997bounds,balke1997probabilistic} provide the first set of inequalities that sharply characterize when the distribution of the observed data is consistent with instrument exogeneity and monotonicity. Their results are obtained through a linear programming formulation. \cite{kitagawa2015test} generalizes these inequalities when the outcome is allowed to be continuous and shows they are sharp constructively. He further proposes a test based on these inequalities. \cite{mourifie2017testing} leverage the intersection bounds framework of \cite{chernozhukov2013intersection} to construct an alternative test based on the same testable implications. When the treatment and the instrument are binary, the model in this paper is equivalent to the model studied in \cite{imbens1994identification}. In that case, we recover the inequalities in \cite{balke1997bounds,balke1997probabilistic} and \cite{kitagawa2015test}. However, both the inequalities and our construction to establish their sharpness beyond this special case are, to our knowledge, novel to the literature.
\cite{kedagni2020generalized} derive generalized instrumental inequalities for a discrete treatment with unrestricted outcomes and instruments, and show that they are sharp when the outcome and treatment are both binary. \cite{kitagawa2021identification} derives a set of sharp inequalities for instrument exogeneity with a continuous outcome, a binary treatment, and a binary instrument. \cite{sun2023instrument} derives a set of inequalities with a possibly multi-valued treatment and instrument, under instrument exogeneity and the ``unordered monotonicity'' assumption of \cite{heckman2018unordered}, but does not establish that these are necessarily sharp. However, the ``unordered monotonicity'' assumption is not implied by, nor does it imply, our assumption. \cite{kwon2026testing} characterize testable implications in the setting with a multi-valued treatment and a binary instrument. Our paper is also related to a vast literature that studies identification and inference for treatment effects using instrumental variables. See, for example, \cite{robins1989analysis}, \cite{cheng2006bounds}, and \cite{song2025categorical}.
Of particular relevance to our setting are the papers which study multi-valued treatments and instruments: see, for instance, \cite{lee2018identifying,lee2026treatment}, \cite{gabriel2024sharp}, \cite{bai2026inference}, and \cite{bai2026identifying}.

The remainder of the paper is organized as follows. In Section \ref{sec:setup}, we describe our setup and notation. In Section \ref{sec:main}, we characterize the set of inequalities implied by our model and show that these are sharp. For simplicity, we first state versions of our results in a setting without an outcome. In Section \ref{sec:extensions}, we extend the results to settings with outcomes and one-sided noncompliance and derive sharp bounds on parameters including the probabilities of substitution patterns. We describe how to test these implications using existing inference methods in Section \ref{sec:inference} and compare our methods to approaches based on latent linear-programming formulation through simulations in Section \ref{sec:sims}. Finally, in Section \ref{sec:empirical}, we apply our methods to the empirical application in \cite{behaghel2013robustness,behaghel2014private}.

\section{Setup and Notation} \label{sec:setup}
Let $J \geq 2$ be an integer and let $D$ denote a multi-valued treatment choice with support $\{0, \dots, J - 1\}$. Let $Z$ denote an instrument with support $\mathcal Z$, where $|\mathcal Z| \geq 2$. In Section \ref{sec:withY}, we additionally consider an outcome variable $Y$, but we ignore it for the time being. In the models that we consider, each value of the instrument encourages at most one unique choice. Let $J_0 \in \{0, \dots, J - 1\}$ denote the number of choices with no encouragement. For convenience, we label the treatment choices and instrument values so that for $J_0 \leq j \leq J - 1$, $Z = j$ encourages $D = j$, while choices $j < J_0$ have no encouragement. In addition, the instrument may have a ``base state'' in which no choices are encouraged. We denote this state by $Z = \mathsf{base}$, which is distinct from $Z = 0, \dots, J - 1$. As a result, $\mathcal Z$ is either $\{J_0, \dots, J - 1\}$\footnote{Because we assume $|\mathcal Z| \geq 2$, $J_0 \leq J - 2$ if $\mathsf{base} \notin \mathcal Z$.} or $\{\mathsf{base}\} \cup \{J_0, \dots, J - 1\}$. Let $D_z$ for $z \in \mathcal Z$ denote the potential treatment choice when assigned to the instrument value $z$. As usual, the observed choice is related to potential treatment choices and the instrument through
\begin{equation} \label{eq:obs}
D = \sum_{z \in \mathcal Z} D_z I \{Z = z\}~.
\end{equation}
Let $Q$ denote the distribution of $((D_z: z \in \mathcal Z), Z)$ and $P$ denote the distribution of $(D, Z)$. Note that given the mapping $T$ such that $(D, Z) = T((D_z: z \in \mathcal Z), Z)$ as implied by \eqref{eq:obs}, we obtain by construction that $P = Q T^{-1}$. Throughout, we will impose the assumption that the instrument is exogenous:
\begin{assumption} \label{as:exog}
$(D_z: z \in \mathcal Z) \indep Z$ under $Q$.
\end{assumption}

Assumption \ref{as:exog} requires that $Z$ is either randomly assigned or ``as good as random.'' See \cite{swanson2018partial} for alternative notions of instrument exogeneity.

In what follows, we will frequently use the following facts about the relationship between $P$ and $Q$. Suppose $P \{Z = z\} > 0$ for all $z \in \mathcal Z$ and $P = Q T^{-1}$ for some $Q$ that satisfies Assumption \ref{as:exog}. Then, for $z \in \mathcal Z$,
\[ P \{Z = z\} = Q \{Z = z\} > 0~.\]
Therefore, the conditional choice probabilities can be defined and they satisfy
\begin{equation} \label{eq:P-Q}
P \{D = j | Z = z\} = Q \{D_z = j | Z = z\} = Q \{D_z = j\}~,
\end{equation}
where the first equality follows because $D = D_z$ when $Z = z$ by \eqref{eq:obs}, and the second equality follows from Assumption \ref{as:exog}.

Our goal is to study the necessary and sufficient conditions for $P$ to be consistent with a class of restrictions on potential treatments that are commonly imposed when analyzing encouragement designs.  Loosely speaking, these restrictions dictate that each value of the instrument encourages at most one unique choice. When $J = 2$ and $\mathcal Z = \{0, 1\}$, we recover the instrument monotonicity condition of \cite{imbens1994identification} that requires $Q \{(D_0, D_1) = (1, 0)\} = 0$. As we explain below, these restrictions, or stronger versions of them, are often used as the key identifying assumptions for the causal interpretation of regression estimands.

\begin{assumption} \label{as:default}
There exists a random variable $\boldsymbol j^{\ast}$ such that $0 \leq \boldsymbol j^{\ast} \leq J - 1$ and 
\begin{equation} \label{eq:default}
Q \{D_j \in \{j, \boldsymbol j^{\ast}\}\} = 1 \text{ for } J_0 \leq j \leq J - 1~.
\end{equation}
Furthermore, if $\mathsf{base} \in \mathcal Z$, then additionally
\begin{equation} \label{eq:default-base}
Q\{\boldsymbol j^{\ast} = D_{\mathsf{base}}\} = 1~.
\end{equation}    
\end{assumption}

To interpret Assumption \ref{as:default}, first suppose $\mathsf{base} \notin \mathcal Z$. Recall that the first $J_0$ choices have no encouragement. Because $Z = j$ encourages $D = j$, it should not affect the comparison among all other choices $\{0, \dots, J - 1\} \setminus \{j\}$.  As a result, if we think of $\boldsymbol j^{\ast}$ as a ``default'' choice (which we note is a random variable since it can differ across individuals) that an individual would have made if the instrument didn't exist, then when $Z = j$, they either stay at $\boldsymbol j^{\ast}$ or are pushed to choose $j$; in particular, they cannot choose any choice that is not $j$ or $\boldsymbol j^{\ast}$. Furthermore, the substitution patterns must be as follows: with a change from $Z = j$ to $Z = k$, they may stay with the original choice if it is the default choice, switch to $k$, or fall back to the default choice $\boldsymbol j^{\ast}$, which may be neither $j$ nor $k$.

Note that Assumption \ref{as:default} rules out a large number of vectors of potential treatments. Consider for instance the setting where $J = 3$ and $\mathcal Z = \{0, 1, 2\}$. The restriction in \eqref{eq:default} implies
\[ Q \{D_0 = 1, D_1 = 2\} = 0~. \]
To see why, suppose $D_0(\omega) = 1$ and $D_1(\omega) = 2$ for some individual $\omega \in \Omega$, where $\Omega$ denotes the underlying probability space. Then, $D_0(\omega) \in \{0, \boldsymbol j^{\ast}(\omega)\}$ implies $\boldsymbol j^{\ast}(\omega) = 1$; at the same time, $D_1(\omega) \in \{1, \boldsymbol j^{\ast}(\omega)\}$ implies $\boldsymbol j^{\ast}(\omega) = 2$, a contradiction. Following similar arguments, we can conclude that under $Q$, $(D_0, D_1, D_2)$ can take at most 10 values with positive probabilities, listed in Table \ref{table:J=3}, instead of $3^3 = 27$ values.

{
\spacingset{\TABLESPACING}
\begin{table}[ht!]
    \centering
    \begin{tabular}{ccccccccccc}
    \toprule
        $D_0$ & 0 & 1 & 2 & 0 & 0 & 1 & 0 & 2 & 0 & 0 \\
        $D_1$ & 0 & 1 & 2 & 1 & 0 & 1 & 1 & 1 & 2 & 1 \\
        $D_2$ & 0 & 1 & 2 & 0 & 2 & 2 & 1 & 2 & 2 & 2 \\
    \bottomrule
    \end{tabular}
    \caption{The support of $(D_0, D_1, D_2)$ if $Q$ satisfies Assumption \ref{as:default} with $\mathcal Z = \{0, 1, 2\}$.}
    \label{table:J=3}
\end{table}
}

\begin{example} \label{eg:behaghel}
\cite{behaghel2013robustness,behaghel2014private} study a randomized controlled trial involving three job search counseling programs in France. In their setting, $Z = 0$ denotes encouragement towards the usual public program without intensive counseling, $Z = 1$ denotes encouragement towards the public program with intensive counseling, and $Z = 2$ denotes encouragement towards the private program. $D = 0, 1, 2$ denotes participation in the corresponding programs. Assumption \ref{as:exog} holds because $Z$ is randomly assigned and we can test if Assumption \ref{as:default} holds using the results in this paper.
\end{example}

\begin{remark} \label{rem:rum}
Assumption \ref{as:default} can be further motivated through a fairly general non-separable random utility model where each instrument increases the utility of at most one unique choice. In that model, $\boldsymbol j^{\ast}$ is the choice that maximizes the baseline utility without the instrument. In Appendix C, we introduce such a model and establish its equivalence to the model consisting of distributions satisfying Assumptions \ref{as:exog} and \ref{as:default}. We note that the same random utility model was shown by \cite{kline2016evaluating} to imply the restrictions in \eqref{eq:kw}. We will in turn show that they are in fact equivalent. 

Our framework is also closely related to the ``universal targeting'' model of Assumption 4 in \cite{lee2026treatment}, which combines what they call ``strict targeting'' and ``one-to-one targeting'' in an additive random utility model. Their Proposition 3 implies our restriction in Assumption \ref{as:default}, with the default selected under a ``non-targeting'' instrument value, which is a base state in our setting. In the special case with $J_0 > 0$ and a base state, ``universal targeting'' imposes the same restrictions on admissible potential treatment vectors as our encouragement-design restriction. Our analysis complements their identification results for treatment effects conditional on complier groups by characterizing the sharp testable implications of encouragement designs, including when outcomes are present and there is no base state.
\end{remark}

When $\mathsf{base} \in \mathcal Z$, the choice under the base state coincides with the default choice, i.e., $\boldsymbol j^{\ast} = D_{\mathsf{base}}$. Then, $Q \{D_j \in \{j, D_{\mathsf{base}}\} \} = 1 \text{ for } J_0 \leq j \leq J - 1$. As illustrated through the examples below, interpreting the assignment to the control group as a base state may be reasonable in some examples but not others, and is in general not a harmless normalization. When $J = 2$, $\mathcal Z = \{0, 1\}$ and $\mathcal Z = \{\mathsf{base}, 1\}$ are equivalent after relabeling $\mathsf{base}$ as $0$; in both cases, Assumption \ref{as:default} simply states $Q \{(D_0, D_1) = (1, 0)\} = 0$, i.e., defiers are ruled out. When $J > 2$, however, interpreting the assignment to control as the base state is no longer without loss of generality. Indeed, consider $J = 3$ as an example. If $\mathcal Z = \{\mathsf{base}, 1, 2\}$, then $Q \{(D_{\mathsf{base}}, D_1, D_2) = (0, 1, 1)\} = 0$, because \eqref{eq:default} and \eqref{eq:default-base} require that $D_2 \in \{D_{\mathsf{base}}, 2\}$ with probability one. If $\mathcal Z = \{0, 1, 2\}$, however, then \eqref{eq:default} allows for $Q \{(D_0, D_1, D_2) = (0, 1, 1)\} > 0$. The reason is that in the latter case, changing $Z = 0$ to $Z = 2$ not only increases the appeal of $D = 2$, but decreases the appeal of $D = 0$ as well (because $D = 0$ is encouraged by $Z = 0$ but not $Z = 2$), making it possible for the person to fall back to the default choice, which in this case is $D = 1$. In contrast, changing $Z = \mathsf{base}$ to $Z = 2$ only increases the appeal of $D = 2$ but does not decrease the appeal of $D = 0$, so one can only switch to $D = 2$ instead of $D = 1$.

To recover a causal interpretation of the IV estimand as a local average treatment effect (LATE) in the setting of Example \ref{eg:behaghel}, \cite{behaghel2013robustness} impose an assumption called ``extended monotonicity.'' As shown in the appendix, this assumption is strictly stronger than the restriction implied by interpreting the assignment to control, $Z = 0$, as a base state. In the empirical application in Section \ref{sec:empirical}, we reject the base-state specification, and thus also their extended monotonicity assumption. Moreover, we use our results to conclude that the substitution pattern $\{D_0 \neq 1, D_2 = 1\}$ is exactly the reason for rejection. At the same time, we cannot reject the model if we view $Z = 0$ as encouraging $D = 0$. The findings therefore suggest that at least for some people, $Z = 0$ strictly increases the appeal of the public program $D = 0$.



\begin{example}\label{eg:kw}
\cite{kline2016evaluating} consider an RCT with a ``close substitute'' to study the effects of preschooling on educational outcomes. In their setting, $D \in \{0,1,2\}$, where $D = 0$ denotes home care (no preschool), $D = 2$ denotes participation in a preschool program called Head Start, and $D = 1$ denotes participation in preschools other than Head Start, namely the close substitute. In this example, $Z \in \{\mathsf{base}, 2\}$, where $Z = 2$ denotes that the household receives an offer to attend Head Start, and $Z = \mathsf{base}$ denotes otherwise. Assumption \ref{as:exog} holds because $Z$ is randomly assigned. In their equation (1), \cite{kline2016evaluating} impose the restriction that
\begin{equation} \label{eq:kw}
Q \{ D_2 = 2 | D_{\mathsf{base}} \neq D_2\}=1~.
\end{equation}

The restriction in \eqref{eq:kw} states that if a household switches their choice upon receiving a Head Start offer, then they must be switching to Head Start. In other words, receiving an offer to Head Start does not change the comparison between no preschool and preschools other than Head Start. The restriction in \eqref{eq:kw} is equivalent to $Q \{D_2 \in \{D_{\mathsf{base}}, 2\}\} = 1$, which is exactly Assumption \ref{as:default} with $J = 3$, and $\mathcal Z = \{\mathsf{base}, 2\}$. The support of $(D_{\mathsf{base}}, D_2)$ is summarized in Table \ref{table:kw}.
{
\spacingset{\TABLESPACING}
\begin{table}[ht!]
    \centering
    \begin{tabular}{cccccc}
    \toprule
    $D_{\mathsf{base}}$ & 0 & 1 & 2 & 0 & 1 \\
    $D_2$ & 0 & 1 & 2 & 2 & 2 \\
    \bottomrule
    \end{tabular}
    \caption{The support of $(D_{\mathsf{base}}, D_2)$ in \cite{kline2016evaluating}.}
    \label{table:kw}
\end{table}
}

The restriction in \eqref{eq:kw} is plausible, although it may be violated because of, for instance, a salience effect, where receiving the Head Start offer makes the household more aware of the merits of preschool in general so they send their children to other preschools.
The key insight behind the empirical results in \cite{kline2016evaluating} is that under the assumption in \eqref{eq:kw}, the Wald estimand from the IV regression identifies a weighted combination of what they call ``sub-LATEs.'' \cite{kline2016evaluating} further establish conditions under which the optimal policy depends upon these ``sub-LATEs.'' Our results in Sections \ref{sec:main} and \ref{sec:withY} will characterize when the distribution of the data is consistent with \eqref{eq:kw}.
\end{example}

In Appendix B, we discuss another important empirical example by \cite{kirkeboen2016field}.


\section{Main Results} \label{sec:main}
\subsection{Sharp Testable Implications} \label{sec:sharp}
In this section, we present our main results on sharp testable implications of Assumptions \ref{as:exog} and \ref{as:default}. Let $\mathbf Q$ denote the set of all distributions of $((D_z: z \in \mathcal Z), Z)$ that satisfy Assumptions \ref{as:exog} and \ref{as:default}. We start with a theorem that characterizes a set of necessary and sufficient conditions for $P$ to be consistent with $\mathbf Q$. We will add an outcome variable to the problem in Section \ref{sec:withY}. To state the theorem, for $0 \leq j \leq J - 1$, define
\[ \mathcal Z(j) = \mathcal Z \setminus \{j\}~. \]
Therefore, for a choice with no encouragement, $\mathcal Z(j) = \mathcal Z$. If $\mathsf{base} \in \mathcal Z$, then $\mathsf{base} \in \mathcal Z(j)$ for all $j$.

\begin{theorem} \label{thm:sharp}
Let $P$ be a probability distribution on $\{0, \dots, J - 1\} \times \mathcal Z$ such that $P \{Z = z\} > 0$ for all $z \in \mathcal Z$. The following results hold:
\begin{enumerate}[\rm (a)]
\item Suppose $P = Q T^{-1}$ for $Q \in \mathbf Q$. Then, for every collection of $z(j) \in \mathcal{Z}(j)$ for $0 \leq j \leq J - 1$, 
\begin{equation} \label{eq:ineq}
\sum_{0 \leq j \leq J - 1} P \{D = j| Z = z(j)\} \leq 1~.
\end{equation}
\item Conversely, suppose \eqref{eq:ineq} holds for every collection of $z(j)\in \mathcal{Z}(j)$ for $0 \leq j \leq J - 1$. Then, there exists a $Q \in \mathbf Q$ such that $P = Q T^{-1}$.
\end{enumerate}
\end{theorem}

Theorem \ref{thm:sharp} contains two parts. Part (a) derives implications of the model $\mathbf Q$ for $P$; the implications in \eqref{eq:ineq} are inequalities in terms of the conditional choice probabilities. Part (b) shows that for each $P$ that satisfies \eqref{eq:ineq}, there exists a distribution $Q \in \mathbf Q$ such that $P = Q T^{-1}$. In other words, the inequalities in Theorem \ref{thm:sharp} are sharp in the sense that they exhaust all the information in the behavioral restrictions imposed by the model $\mathbf Q$.

We first explain why Theorem \ref{thm:sharp}(a) holds, and then discuss sharpness in Section \ref{sec:construction}. The inequalities in \eqref{eq:ineq} are direct consequences of the restriction in \eqref{eq:default}. To see why, fix $z(j) \in \mathcal Z(j)$ for $0 \leq j \leq J - 1$ and consider the events
\[ \{D_{z(0)} = 0\}, \dots, \{D_{z(J - 1)} = J - 1\}~. \]
For $0 \leq j \leq J - 1$, $D_{z(j)} = j$ implies $\boldsymbol j^{\ast} = j$. If $z(j) \neq \mathsf{base}$, then this follows from \eqref{eq:default} and $z(j) \neq j$; if $z(j) = \mathsf{base}$, then this follows from \eqref{eq:default-base}. As a result, the events listed above are disjoint, so the sum of their probabilities is at most one, and \eqref{eq:ineq} follows from Assumption \ref{as:exog}.

The inequalities in \eqref{eq:ineq} restrict the substitution patterns jointly at different values of the instrument $z(0), \dots, z(J - 1)$. In particular, it does not suffice to consider pairwise substitution patterns when the instrument changes from one value to another. Note that $z(0), \dots, z(J - 1)$ do not have to be distinct. Therefore, for $J_0 \leq j \leq J - 1$ and $z \in \mathcal Z(j)$, by setting $z(j) = z$ and $z(\ell) = j$ for all $\ell \neq j$, we have
\[ P \{D = j | Z = z\} + \sum_{\ell \neq j} P \{D = \ell | Z = j\} \leq 1~, \]
which implies
\begin{equation} \label{eq:encourage}
    P \{D = j | Z = z\} \leq P \{D = j | Z = j\}~.
\end{equation}
The inequality in \eqref{eq:encourage} states that the conditional probability of choosing $j$ is maximized at $Z = j$, which aligns with the intuition that $Z = j$ encourages choice $D = j$. When $J = 2$ and $\mathcal Z = \{0, 1\}$, we recover the well-known inequality $P \{D = 1 | Z = 0\} \leq P \{D = 1 | Z = 1\}$ implied by the instrument monotonicity condition in \cite{imbens1994identification}.

\begin{example}
Suppose $\mathcal Z = \{0, \dots, J - 1\}$. When $J = 2$, Theorem \ref{thm:sharp} produces only one inequality:
\[ P \{D = 1 | Z = 0\} \leq P \{D = 1 | Z = 1\}~. \]
When $J = 3$, \eqref{eq:encourage} leads to six inequalities. In addition, by considering the cases where $z(0), z(1), z(2)$ are all distinct in \eqref{eq:ineq}, we have
\begin{align*}
P \{D = 1 | Z = 0\} + P \{D = 2 | Z = 1\} + P \{D = 0 | Z = 2\} & \leq 1 \\
P \{D = 2 | Z = 0\} + P \{D = 0 | Z = 1\} + P \{D = 1 | Z = 2\} & \leq 1~.
\end{align*}
These two inequalities involve the substitution patterns across \emph{triplets} of values of the instrument instead of simple pairwise comparison. In total, we obtain eight inequalities. For a general $J$, we obtain $(J - 1)^J$ inequalities.
\end{example}

When $\mathsf{base} \in \mathcal Z$, we obtain the following simplification of Theorem \ref{thm:sharp}:

\begin{corollary} \label{cor:0better}
Let $P$ be a probability distribution on $\{0, \dots, J - 1\} \times \mathcal Z$ such that $P \{Z = z\} > 0$ for all $z \in \mathcal Z$. Suppose $\mathsf{base} \in \mathcal Z$. Then, the inequalities described by \eqref{eq:ineq} are equivalent to the statement that, for $0 \leq j \leq J - 1$ and $z \in \mathcal Z(j)$,
\begin{equation} \label{eq:0better}
P \{D = j | Z = z\} \leq P \{D = j | Z = \mathsf{base}\}~. 
\end{equation}
\end{corollary}

The inequality in \eqref{eq:0better} follows immediately from Assumption \ref{as:default} when $\mathsf{base} \in \mathcal Z$. It is trivial if $z = \mathsf{base}$. If $z \in \mathcal Z \setminus \{\mathsf{base}, j\}$ and $D_z = j$, then because \eqref{eq:default} and \eqref{eq:default-base} imply $D_z \in \{z, D_{\mathsf{base}}\}$, it has to be the case that $D_{\mathsf{base}} = j$. Therefore,
\[ \{D_z = j\} \implies \{D_{\mathsf{base}} = j\}~, \]
which immediately implies \eqref{eq:0better}. Because these inequalities are sharp, we conclude that when $\mathsf{base} \in \mathcal Z$, all information in the data about whether the data is consistent with the model is contained in the \emph{pairwise} comparison between the choices when $Z = \mathsf{base}$ versus $Z = z$. Therefore, unlike when $\mathsf{base} \notin \mathcal Z$, here the substitution patterns between $Z = k$ and $Z = \ell$ for $k, \ell \neq \mathsf{base}$ and $k \neq \ell$ do not contain additional information.

We now apply the results in Corollary \ref{cor:0better} to the examples.

\begin{example}
Recall in Example \ref{eg:kw} that $\mathcal Z = \{\mathsf{base}, 2\}$. In this case, two inequalities follow from \eqref{eq:0better}:
\begin{align*}
P \{D = 0 | Z = 2\} & \leq P \{D = 0 | Z = \mathsf{base}\} \\
P \{D = 1 | Z = 2\} & \leq P \{D = 1 | Z = \mathsf{base}\}~.
\end{align*}
There is no additional inequality for $j = 2$.
\end{example}

\begin{example}
If $\mathcal Z = \{\mathsf{base}, 1, 2\}$, then \eqref{eq:0better} produces the following four inequalities:
\begin{align*}
P \{D = 0 | Z = 1\} & \leq P \{D = 0 | Z = \mathsf{base}\} \\
P \{D = 0 | Z = 2\} & \leq P \{D = 0 | Z = \mathsf{base}\} \\
P \{D = 1 | Z = 2\} & \leq P \{D = 1 | Z = \mathsf{base}\} \\
P \{D = 2 | Z = 1\} & \leq P \{D = 2 | Z = \mathsf{base}\}~.
\end{align*}
In an empirical application to the dataset for \cite{behaghel2013robustness,behaghel2014private} in Section \ref{sec:empirical}, these inequalities and their counterparts with an outcome will be used to test the specification that interprets the assignment to control as a base state.
\end{example}


\begin{remark} \label{rem:implicit}
Following \cite{balke1997probabilistic}, \cite{gabriel2024sharp}, and \cite{bai2026inference}, it can be shown that there exists $Q \in \mathbf Q$ such that $P = Q T^{-1}$ if and only if a certain linear system has a nonnegative solution.
This approach provides an \emph{implicit} characterization of the problem. Based on this observation, one could also construct a test that $P$ is consistent with $\mathbf Q$ using projection-type inference methods for linear moment inequalities. In Section \ref{sec:sims} we compare such methods to the tests we develop in Section \ref{sec:inference} using our closed-form characterization.
\end{remark}

\begin{remark} \label{rem:baietal}
Suppose $\mathcal Z = \{0, \dots, J - 1\}$. As noted in Remark \ref{rem:rum}, $\mathbf Q$ is equivalent to a general non-separable random utility model. We emphasize, however, that Assumption \ref{as:default} is stronger than the generalized monotonicity restriction considered in \cite{bai2026identifying}, which in the current context states
\begin{equation} \label{eq:baietal}
Q \{D_j \neq j, D_k = j \text{ for some } k \neq j\} = 0~.
\end{equation}
The restriction \eqref{eq:baietal} can be shown to be equivalent to \eqref{eq:default} when $J = 3$, but is weaker than \eqref{eq:default} when $J \geq 4$. Indeed, when $J = 4$, $(D_0, D_1, D_2, D_3) = (1, 1, 2, 2)$ is not ruled out by \eqref{eq:baietal} but is ruled out by \eqref{eq:default}, because $D_0(\omega) = 1 \neq 0$ implies $\boldsymbol j^{\ast}(\omega) = 1$, whereas $D_3(\omega) = 2 \neq 3$ implies $\boldsymbol j^{\ast}(\omega) = 2$, a contradiction. Moreover, for $P = Q T^{-1}$ where $Q$ satisfies Assumption \ref{as:exog} and $Q \{(D_0, D_1, D_2, D_3) = (1, 1, 2, 2)\} = 1$, we have $P \{D = 1 | Z = 0\} + P \{D = 2 | Z = 3\} = 2$, violating \eqref{eq:ineq}. This example shows that the stronger restrictions imposed by $\mathbf Q$ generate strictly stronger testable implications.
\end{remark}

\subsection{Constructive Proof of Sharpness} \label{sec:construction}
Theorem \ref{thm:sharp}(b) shows that the inequalities in \eqref{eq:ineq} are sharp in the sense that for each $P$ that satisfies \eqref{eq:ineq}, there exists a distribution $Q \in \mathbf Q$ such that $P = Q T^{-1}$. In the special case of $J = 2$ and $\mathcal Z = \{0, 1\}$, a constructive proof was provided by \cite{kitagawa2015test}, but the proof does not extend to the case when $J > 2$. In particular, the construction in \cite{kitagawa2015test} relies crucially on the observation that if $D = 1$ when $Z = 0$, then $(D_0, D_1) = (1, 1)$; this subgroup of people are referred to as the ``always takers,'' and the group that takes $D = 0$ when $Z = 1$ are called the ``never takers.'' The remaining probability mass is then assigned to the ``compliers,'' for whom $(D_0, D_1) = (0, 1)$. When $J > 2$, however, it is in general impossible to pin down the joint distribution of potential treatments using this argument, and hence the proof requires an entirely new strategy. Importantly, the proof that we present is still constructive, in that we construct $Q$ explicitly from the given distribution $P$.

To illustrate why \eqref{eq:ineq} is sufficient for determining whether $P$ is consistent with the model $\mathbf Q$, we sketch the construction when $J = 3$ and $\mathcal Z = \{0, 1, 2\}$. Let $Q^\ast$ denote a candidate distribution for which we wish to show $P = Q^\ast T^{-1}$ and $Q^\ast \in \mathbf Q$. We start by considering $P \{D = 0 | Z = z\}$. First note that if $P = Q^\ast T^{-1}$, then for $z \in \{0, 1, 2\}$,
\[ P \{D = 0 | Z = z\} = Q^\ast \{D_z = 0\} \geq Q^\ast \{(D_0, D_1, D_2) = (0, 0, 0)\}~. \]
Respecting this constraint, we set
\[ Q^\ast \{(D_0, D_1, D_2) = (0, 0, 0)\} = \min_{z \in \mathcal Z} P \{D = 0 | Z = z\}~. \]
The inequalities in \eqref{eq:encourage} imply the minimum on the right-hand side is attained at $z \in \{1, 2\}$, and without loss of generality suppose it is attained at $z = 1$. If it is attained at $z = 2$, then a symmetric construction applies. Next, note any candidate $Q^\ast$ has to satisfy
\begin{align*}
    P \{D = 0 | Z = 2\} & = Q^\ast \{D_2 = 0\} \\
    & = Q^\ast \{(D_0, D_1, D_2) = (0, 0, 0)\} + Q^\ast \{(D_0, D_1, D_2) = (0, 1, 0)\} \\
    & = P \{D = 0 | Z = 1\} + Q^\ast \{(D_0, D_1, D_2) = (0, 1, 0)\}~,
\end{align*}
and hence we have to define
\[ Q^\ast \{(D_0, D_1, D_2) = (0, 1, 0)\} = P \{D = 0 | Z = 2\} - P \{D = 0 | Z = 1\}~. \]
This step stops here.
By construction,
\[ Q^\ast \{D_z = 0\} = P \{D = 0 | Z = z\} \text{ for } z \neq 0~. \]
The total mass we have assigned in this step is
\begin{align*}
    Q^\ast \{(D_0, D_1, D_2) = (0, 0, 0)\} + Q^\ast \{(D_0, D_1, D_2) = (0, 1, 0)\} = \max_{z \neq 0} P \{D = 0 | Z = z\}~.
\end{align*}

Similarly, we can carry out the construction for events corresponding to $P \{D = 1 | Z = z\}$ and $P \{D = 2 | Z = z\}$. The total mass assigned after these steps is
\[ \sum_{0 \leq j \leq 2} \max_{z(j) \neq j} P \{D = j | Z = z(j)\} \leq 1~, \]
where the inequality is precisely \eqref{eq:ineq}. We then assign the remaining mass to the diagonal:
\[ Q^\ast \{(D_0, D_1, D_2) = (0, 1, 2)\} = 1 - \sum_{0 \leq j \leq 2} \max_{z(j) \neq j} P \{D = j | Z = z(j)\} \geq 0~. \]
Because in the construction we have guaranteed $Q^\ast \{D_z = j\} =  P \{D = j | Z = z\}$ for $j \neq z$, it is straightforward to verify the equality also for $j = z$, and hence $P = Q^\ast T^{-1}$.

\section{Extensions} \label{sec:extensions}
\subsection{Results with Outcomes} \label{sec:withY}
In this subsection, we present the general results with an outcome variable. The discussion runs mostly in parallel with Section \ref{sec:main}. Let $\mathcal Y$ be a Borel subset of a Euclidean space. Let $Y \in \mathcal Y$ denote the observed outcome and $Y_d \in \mathcal Y$ for $0 \leq d \leq J - 1$ denote the potential outcome under treatment choice $d$. Note that $Y$ can be vector-valued and we put no restriction on its support beyond it being a Borel set. In addition to \eqref{eq:obs}, the observed outcome and the potential outcomes are related through
\[ Y = \sum_{0 \leq d \leq J - 1} Y_d I \{D = d\}~. \]
Note that we have assumed an individual-level exclusion restriction in the sense that for each individual, the instrument does not enter the potential outcome directly. With some abuse of notation, we continue to let $T(\cdot)$ denote the mapping defined by the equation above together with \eqref{eq:obs}. Let $P$ denote the distribution of $(Y, D, Z)$ and $Q$ denote the distribution of $(Y_0, \dots, Y_{J - 1}, (D_z: z \in \mathcal Z), Z)$. We modify Assumption \ref{as:exog} to include the potential outcomes:
\begin{assumption} \label{as:exog-Y}
$(Y_0, \dots, Y_{J - 1}, (D_z: z \in \mathcal Z)) \indep Z$ under $Q$.
\end{assumption}
Let $\mathbf Q^{\mathsf Y}$ denote the set of all distributions $Q$ for which Assumptions \ref{as:exog-Y} and \ref{as:default} hold. We first present the counterpart to Theorem \ref{thm:sharp}.

\begin{theorem} \label{thm:sharp-Y}
Let $P$ be a probability distribution on $\mathcal Y \times \{0, \dots, J - 1\} \times \mathcal Z$ such that $P \{Z = z\} > 0$ for all $z \in \mathcal Z$. Then, $P = Q T^{-1}$ for $Q \in \mathbf Q^{\mathsf Y}$ if and only if both of the following sets of conditions hold:
\begin{enumerate}[\rm (a)]
\item For every collection of Borel sets $\{B_z(j): z \in \mathcal Z(j)\}$, $0 \leq j \leq J - 1$, such that $\{B_z(j): z \in \mathcal Z(j)\}$ forms a partition of $\mathcal Y$ for each $j$, we have
\begin{equation} \label{eq:ineq-Y}
\sum_{0 \leq j \leq J - 1} \sum_{z \in \mathcal Z(j)} P \{Y \in B_z(j), D = j | Z = z\} \leq 1~.
\end{equation}
\item For every Borel set $B \subseteq \mathcal Y$, $J_0 \leq j \leq J - 1$, and $z \in \mathcal Z(j)$,
\begin{equation} \label{eq:encourage-Y}
P \{Y \in B, D = j | Z = z\} \leq P \{Y \in B, D = j | Z = j\}~.
\end{equation}
\end{enumerate}
\end{theorem}

For $z(j) \in \mathcal Z(j)$, $0 \leq j \leq J - 1$, note that by taking $B_{z(j)}(j) = \mathcal Y$ and $B_z(j) = \emptyset$ for $0 \leq j \leq J - 1$ and $z \neq z(j)$ in \eqref{eq:ineq-Y}, we recover \eqref{eq:ineq} in Theorem \ref{thm:sharp}.

When $\mathsf{base} \in \mathcal Z$, as in Corollary \ref{cor:0better}, we obtain the following simplification of Theorem \ref{thm:sharp-Y}.

\begin{corollary} \label{cor:0better-Y}
Let $P$ be a probability distribution on $\mathcal Y \times \{0, \dots, J - 1\} \times \mathcal Z$ such that $P \{Z = z\} > 0$ for all $z \in \mathcal Z$. Suppose $\mathsf{base} \in \mathcal Z$. Then, $P \in \mathbf Q^{\mathsf Y} T^{-1}$ if and only if the conditions in Theorem \ref{thm:sharp-Y}(b) hold and for every Borel set $B \subseteq \mathcal Y$, $0 \leq j \leq J - 1$ and $z \in \mathcal Z(j)$,
\begin{equation} \label{eq:0better-Y}
P \{Y \in B, D = j | Z = z\} \leq P \{Y \in B, D = j | Z = \mathsf{base}\}~.
\end{equation}
\end{corollary}

\begin{remark}
When $J = 2$ and $\mathcal Z = \{0, 1\}$, the only inequalities implied by Theorem \ref{thm:sharp-Y} are
\begin{align*}
    P \{Y \in B, D = 1 | Z = 0\} & \leq P \{Y \in B, D = 1 | Z = 1\} \\
    P \{Y \in B, D = 0 | Z = 1\} & \leq P \{Y \in B, D = 0 | Z = 0\}~.
\end{align*}
Note that these inequalities coincide with the simplification described in Corollary \ref{cor:0better-Y} for $J = 2$ and $\mathcal Z = \{\mathsf{base}, 1\}$ if we relabel $\mathsf{base}$ as $0$. Furthermore, these inequalities are exactly those derived by \cite{balke1997bounds,balke1997probabilistic} and \cite{kitagawa2015test}. We emphasize, however, that both the inequalities and the arguments to establish sharpness are novel beyond this setting. 
\end{remark}

\subsection{Sharp Bounds on Substitution Patterns}
The construction in the proofs of Theorems \ref{thm:sharp} and \ref{thm:sharp-Y} also leads to sharp lower bounds on the probabilities of many substitution patterns. In particular, the same distribution constructed in the proof of Theorem \ref{thm:sharp-Y} attains the lower bounds of these parameters simultaneously.

\begin{theorem} \label{thm:subs}
Let $P$ be a probability distribution on $\mathcal Y \times \{0, \dots, J - 1\} \times \mathcal Z$ such that $P \{Z = z\} > 0$ for all $z \in \mathcal Z$ and suppose there exists $Q \in \mathbf Q^{\mathsf Y}$ such that $P = Q T^{-1}$. Then, for $0 \leq j \leq J - 1$, $z, z' \in \mathcal Z$, $z \neq z'$, and every Borel set $B \subseteq \mathcal Y$, we have
\begin{multline} \label{eq:subs}
\min_{Q \in \mathbf Q^{\mathsf Y}: P = Q T^{-1}} Q \{Y_j \in B, D_z = j, D_{z'} \neq j\} \\
= \sup_{C \subseteq B} \big ( P \{Y \in C, D = j | Z = z\} - P \{Y \in C, D = j | Z = z'\} \big )~,
\end{multline}
where the supremum is over Borel sets $C \subseteq B$. Moreover, the distribution constructed in the proof of Theorem \ref{thm:sharp-Y} attains these minima simultaneously for every Borel set $B$ and all $j$, $z$, and $z'$ as stated.
\end{theorem}

Taking $B = \mathcal Y$ in \eqref{eq:subs}, we get the sharp lower bound on $Q \{D_z = j, D_{z'} \neq j\}$. In particular, if $\mathcal Y$ is finite, then the lower bound is
\begin{equation} \label{eq:subs-finite}
\sum_{y \in \mathcal Y} \big ( P \{Y = y, D = j | Z = z\} - P \{Y = y, D = j | Z = z'\} \big )_+~,
\end{equation}
where $x_+ = \max \{x, 0\}$.

Considering all different values of $j$, Theorem \ref{thm:subs} also sharply characterizes the minimum fraction of people whose treatment must change between any two instrument values. In establishing the following result, simultaneous attainment in Theorem \ref{thm:subs} is crucial, because in general the sharp lower bound of the sum of parameters need not equal the sum of their sharp lower bounds. Moreover, these lower bounds are attained for all pairs of distinct instruments simultaneously. For probability measures $\mu$ and $\mu'$ on the same measurable space, let $\|\mu - \mu'\|_{\mathrm{TV}} = \sup_A |\mu(A) - \mu'(A)|$
denote their total variation distance.

\begin{corollary} \label{cor:tv}
Suppose the conditions of Theorem \ref{thm:subs} hold. Then, for $z, z' \in \mathcal Z$ such that $z \neq z'$,
\begin{equation} \label{eq:tv-switching}
\min_{Q \in \mathbf Q^{\mathsf Y}: P = Q T^{-1}} Q \{D_z \neq D_{z'}\} = \|P\{(Y,D) \in \cdot | Z = z\} - P\{(Y,D) \in \cdot | Z = z'\}\|_{\mathrm{TV}}~.
\end{equation}
Moreover, the distribution constructed in the proof of Theorem \ref{thm:sharp-Y} attains these minima simultaneously for all distinct $z, z' \in \mathcal Z$.
\end{corollary}

Suppose $k, \ell \in \{J_0, \dots, J - 1\}$ and $k \neq \ell$. When the instrument switches from $\ell$ to $k$, some people may switch into $D = k$, given by the event $\{D_\ell \neq k, D_k = k\}$. Theorem \ref{thm:subs} implies the following corollary that allows us to decompose the source of these people according to the values of $D_\ell$. It implies sharp lower bounds on the probabilities $Q\{D_\ell = j, D_k = k\}$ for $j \notin \{k, \ell\}$ and the sharp upper bound on the probability of ``pairwise compliers'' defined by $\{D_\ell = \ell, D_k = k\}$. Moreover, these sharp bounds hold simultaneously because they are attained by the same distribution constructed in the proof of Theorem \ref{thm:sharp-Y}.

\begin{corollary} \label{cor:sources}
Let $P$ be a probability distribution on $\mathcal Y \times \{0, \dots, J - 1\} \times \mathcal Z$ such that $P \{Z = z\} > 0$ for all $z \in \mathcal Z$. Suppose there exists $Q \in \mathbf Q^{\mathsf Y}$ such that $P = Q T^{-1}$. Suppose $k, \ell \in \{J_0, \dots, J - 1\}$ and $k \neq \ell$. Then, for all such $Q$,
\begin{equation} \label{eq:entry}
Q \{D_\ell \neq k, D_k = k\} = P \{D = k | Z = k\} - P \{D = k | Z = \ell\}~.
\end{equation}
For $0 \leq j \leq J - 1$ with $j \notin \{k,\ell\}$ and every Borel set $B \subseteq \mathcal Y$,
\begin{multline} \label{eq:source}
\min_{Q \in \mathbf Q^{\mathsf Y}: P = Q T^{-1}} Q \{Y_j \in B, D_\ell = j, D_k = k\} \\
= \sup_{C \subseteq B} \big ( P \{Y \in C, D = j | Z = \ell\} - P \{Y \in C, D = j | Z = k\} \big )~,
\end{multline}
where the supremum is over Borel sets $C \subseteq B$. Moreover,
\begin{equation} \label{eq:pairwise-complier}
\begin{aligned} 
& \max_{Q \in \mathbf Q^{\mathsf Y}: P = Q T^{-1}} Q \{D_\ell = \ell, D_k = k\} \\
&  = P \{D = k | Z = k\} - P \{D = k | Z = \ell\} \\
& \hspace{2em} - \sum_{0 \leq j \leq J - 1, j \notin \{k,\ell\}} \sup_{B \subseteq \mathcal Y} \big ( P \{Y \in B, D = j | Z = \ell\} - P \{Y \in B, D = j | Z = k\} \big )~,
\end{aligned}
\end{equation}
where each supremum is over Borel sets $B \subseteq \mathcal Y$. Moreover, the distribution constructed in the proof of Theorem \ref{thm:sharp-Y} attains the optima in \eqref{eq:source} and \eqref{eq:pairwise-complier} simultaneously for all Borel sets and indices as stated.
\end{corollary}

\begin{remark}
The sharp bounds in this subsection can be unified using submodular functions. For each $0 \leq j \leq J - 1$, let $\mathcal W_j = \{z \in \mathcal Z : D_z = j\}$ denote the set of instrument values under which an individual chooses treatment $j$. Recall that a function $f:2^{\mathcal Z} \to \mathbf R$ is submodular if
\[ f(S) + f(S') \geq f(S \cap S') + f(S \cup S') \]
for all $S, S' \subseteq \mathcal Z$. For each $0 \leq j \leq J - 1$, let $\varphi_j: \mathcal Y \times 2^{\mathcal Z} \to \mathbf R$ be bounded and Borel measurable in its first argument for every $S \subseteq \mathcal Z$, and suppose $\varphi_j(y, \emptyset) = 0$ and $\varphi_j(y, \cdot)$ is submodular for every $y \in \mathcal Y$. Then, one can establish the sharp lower bound on $\sum_{0 \leq j \leq J - 1} E_Q[\varphi_j(Y_j, \mathcal W_j)]$.
Moreover, the same distribution constructed in the proof of Theorem \ref{thm:sharp-Y} attains this sharp lower bound simultaneously for all collections of functions satisfying the conditions above. The sharp bounds in Theorem \ref{thm:subs} and Corollaries \ref{cor:tv} and \ref{cor:sources} can be recovered as special cases.
\end{remark}

\subsection{One-Sided Noncompliance} \label{sec:onesided}
In this section, we extend our results to settings with one-sided noncompliance. Formally, let $\mathcal Z = \{\mathsf{base}, 1, \dots, J - 1\}$ and $\mathbf Q_{\mathsf{1s}}^{\mathsf Y}$ denote the collection of all distributions $Q$ of $(Y_0, \dots, Y_{J - 1}, D_{\mathsf{base}}, D_1, \dots, D_{J - 1}, Z)$ such that Assumption \ref{as:exog-Y} holds and
\begin{equation} \label{eq:onesided}
Q \{D_{\mathsf{base}} = 0\} = 1 \text{ and } Q \{D_j \in \{j, 0\}\} = 1 \text{ for } 1 \leq j \leq J - 1~.
\end{equation}
The condition in \eqref{eq:onesided} requires that when assigned $Z = j$, the person either takes up $D = j$ or the control status $D = 0$. Therefore, noncompliance can only be one-sided ($j$ to $0$) instead of the other way around. The only difference between \eqref{eq:onesided} and Assumption \ref{as:default} when $\mathcal Z = \{\mathsf{base}, 1, \dots, J - 1\}$ is that we additionally require $D_{\mathsf{base}} = 0$. Such a setting is prevalent in practice, especially if the instrument is the eligibility for each program, so that one either takes up the program they are eligible for or falls back to control. See \cite{kirkeboen2016field} and \cite{angrist2009incentives} for examples. The following theorem presents the sharp testable implications of \eqref{eq:onesided}.

\begin{theorem} \label{thm:onesided}
Let $P$ be a probability distribution on $\mathcal Y \times \{0, \dots, J - 1\} \times \mathcal Z$ such that $P \{Z = z\} > 0$ for all $z \in \mathcal Z$. Then, $P = Q T^{-1}$ for some $Q \in \mathbf Q_{\mathsf{1s}}^{\mathsf Y}$ if and only if $P \{D = 0 | Z = \mathsf{base}\} = 1$, $P \{D = j | Z = z\} = 0$ for $z \in \{1, \dots, J - 1\}$ and $j \notin \{z, 0\}$, and for each Borel set $B \subseteq \mathcal Y$ and $1 \leq j \leq J - 1$,
\begin{equation} \label{eq:onesided-testable}
P \{Y \in B, D = 0 | Z = j\} \leq P \{Y \in B, D = 0 | Z = \mathsf{base}\}~.
\end{equation}
\end{theorem}

We note that the results in Theorem \ref{thm:onesided} follow from Corollary \ref{cor:0better-Y} once we impose that $P \{D = 0 | Z = \mathsf{base}\} = 1$ and $P \{D = j | Z = z\} = 0$ for $z \in \{1, \dots, J - 1\}$ and $j \notin \{z, 0\}$. Indeed, with this additional restriction, \eqref{eq:encourage-Y} is vacuous because the left-hand side is always 0. At the same time, both sides of \eqref{eq:0better-Y} are 0 for $j \neq 0$, so \eqref{eq:0better-Y} is only meaningful when $j = 0$, becoming \eqref{eq:onesided-testable}.

\section{Inference} \label{sec:inference}
In this section, using the characterizations in Theorem \ref{thm:sharp-Y} or Corollary \ref{cor:0better-Y}, we develop tests to assess if the model is consistent with the distribution of the data. Formally, let $\mathbf P$ denote some fixed class of distributions and $\mathbf P_0 = \{P \in \mathbf P: \text{ there is } Q \in \mathbf Q^{\mathsf Y} \text{ such that } P = Q T^{-1}\}$. Given an i.i.d.\ sample $(Y_i, D_i, Z_i)$, $1 \leq i \leq n$ from $P \in \mathbf P$, we test
\begin{equation} \label{eq:null}
H_0: P \in \mathbf P_0 \text{ versus } H_1: P \in \mathbf P \setminus \mathbf P_0
\end{equation}
in a way that uniformly controls size across $\mathbf P_0$.

The conditions in Theorem \ref{thm:sharp-Y}, Corollary \ref{cor:0better-Y}, and Theorem \ref{thm:onesided} require checking inequalities for every Borel set or every collection of Borel sets that form a partition of $\mathcal Y$. The following lemma allows us to focus on cells instead. Let $\mathrm{diam}(\cdot)$ denote the diameter of a set in a Euclidean space.

\begin{lemma} \label{lem:cells}
For each $m \geq 1$, let $\mathcal I_m$ be a partition of $\mathcal Y$ consisting of a finite number of Borel subsets of $\mathcal Y$. Suppose for each $m \geq 1$, $\mathcal I_{m + 1}$ is a refinement of $\mathcal I_m$. Further suppose for each $y \in \mathcal Y$, the set in $\mathcal I_m$ containing $y$, denoted by $I_m(y)$, satisfies $\mathrm{diam}(I_m(y)) \to 0$. Then, the inequalities in \eqref{eq:ineq-Y} hold if and only if for every $m \geq 1$, for every collection of $z(j, I) \in \mathcal Z(j)$ for $0 \leq j \leq J - 1$ and $I \in \mathcal I_m$,
\begin{equation} \label{eq:ineq-cells}
\sum_{0 \leq j \leq J - 1} \sum_{I \in \mathcal I_m} P \{Y \in I, D = j | Z = z(j, I)\} \leq 1~.
\end{equation}
In addition, the inequalities in \eqref{eq:encourage-Y}, \eqref{eq:0better-Y}, and \eqref{eq:onesided-testable} hold for every Borel set $B \subseteq \mathcal Y$ if and only if, respectively, for every $m \geq 1$ and $I \in \mathcal I_m$,
\begin{align}
\label{eq:ineq-cells-2} P \{Y \in I, D = j | Z = z\} & \leq P \{Y \in I, D = j | Z = j\} \\
\label{eq:ineq-cells-3} P \{Y \in I, D = j | Z = z\} & \leq P \{Y \in I, D = j | Z = \mathsf{base}\} \\
\label{eq:ineq-cells-4} P \{Y \in I, D = 0 | Z = j\} & \leq P \{Y \in I, D = 0 | Z = \mathsf{base}\}~.
\end{align}
\end{lemma}


Using similar arguments to those in the proof of Lemma \ref{lem:cells}, it is straightforward to show that the same partitions also approximate the sharp bounds in Theorem \ref{thm:subs} and Corollaries \ref{cor:tv} and \ref{cor:sources}. When $\mathcal Y$ is finite, Lemma \ref{lem:cells} further simplifies by taking $\mathcal I_m$ to be the collection of singletons of $\mathcal Y$. In any case, for each fixed $m \geq 1$, Lemma \ref{lem:cells} implies a finite collection of moment inequalities $\gamma(P) \geq 0$, which can be tested using existing inference methods for moment inequalities \citep[see, for instance,][for an overview]{pakes2017practical}. In contrast, constructing tests based on an implicit characterization of the problem, as discussed in Remark \ref{rem:implicit}, in practice requires methods designed for projection inference in linear systems, such as \cite{fang2023inference}. In Section \ref{sec:sims}, we compare the power and computation time of both formulations using \cite{fang2023inference} as a common inference procedure.

A further benefit of the constructive characterization given in Lemma \ref{lem:cells} is that it facilitates multiple hypothesis testing of the individual choice inequalities via, for instance, the step-down procedure developed in \cite{romano2018multiple}. In Section \ref{sec:empirical}, we perform such an exercise on the dataset from \cite{behaghel2013robustness} to detect which substitution patterns are responsible for rejection of the base-state specification.



\section{Simulations} \label{sec:sims}
In this section, we study the properties of the tests described in Section \ref{sec:inference} for the null hypothesis in \eqref{eq:null}. We consider both finite and continuous outcomes and compare two approaches for inference. To compare both approaches using a common inference procedure, we apply the test in \cite{fang2023inference} to both, although we note that the first approach could use existing methods for moment inequalities other than those designed for projection inference in linear systems. The first approach tests our sharp inequalities in closed form. When the outcome is continuous, we test the inequalities in Lemma \ref{lem:cells} for different levels of discretization. Formally, testing the inequalities $\gamma(P) \geq 0$ is equivalent to determining the existence of $x \geq 0$ such that $A x = \gamma(P)$, where $A$ is the identity matrix. The second approach tests if there exists $x \geq 0$ such that $A^{\sf latent} x = \beta^{\sf latent}(P)$, where for the finite outcome or a given level of discretization of the continuous outcome, $\beta^{\sf latent}(P)$ collects the conditional probabilities of $Y$ and $D$ given $Z$, $x$ represents the latent probabilities under $Q$ that satisfies Assumption \ref{as:default}, and $A^{\sf latent}$ is a matrix defined through $P \{Y \in B, D = d | Z = z\} = Q \{Y_d \in B, D_z = d\}$. For finite outcomes, the null hypothesis holds if and only if this linear program is feasible, as shown in \cite{balke1997probabilistic} and \cite{bai2026inference}. For continuous outcomes, we follow \cite{bai2026inference} to test linear programs for discretized outcomes. We refer to the two approaches as the ``Closed Form'' and ``Latent'' approaches.

We implement both approaches using the \texttt{lpinfer} package in \textsf{R} and the same tuning parameters for each specification and outcome partition.\footnote{Let $K$ denote the number of outcome categories, with $K = 1$ for tests without an outcome, and let $p_K$ be the larger row count of the two systems, including known rows and before removing redundant equations. For both approaches, we set $\lambda_{n, K} = \{\log(e \vee p_K) \log(e \vee \log(e \vee n))\}^{-1/2}$ and $\rho = 10^{-4}$.} We consider sample sizes $n \in \{500, 1000, 2000\}$ and report rejection rates in percentages for tests at the 5\% level based on $1{,}000$ replications and $500$ bootstrap draws in each replication. Within each specification and sample size, we use the same underlying random draws across tests, outcome partitions, and values of $\theta$. The tables also share bootstrap draws across these comparisons. The simulations are computed using 18 parallel workers and one solver thread per worker on an Apple MacBook Pro with an 18-core M5 Pro processor and 48 GB of unified memory.

\subsection{Finite Outcomes} \label{sec:sims-finite}
We begin by describing the general setup of the distributions used in the simulations and then specialize to a finite outcome. Without a base state, we set $J = 3$ and $\mathcal Z = \{0, 1, 2\}$. With a base state, we set $J = 4$ and $\mathcal Z = \{\mathsf{base}, 1, 2, 3\}$. To describe both specifications together, let $z_0 = 0$ in the first and $z_0 = \mathsf{base}$ in the second. We draw $Z$ uniformly from $\mathcal Z$. Independently of $Z$, we draw $D_{z_0}$ uniformly from $\{0, \dots, J - 1\}$ and $C_j \sim \mathrm{Bernoulli}(1/4)$ for $1 \leq j \leq J - 1$, with all these draws independent. We then set $D_j = j$ if $C_j = 1$ and $D_j = D_{z_0}$ otherwise. The observed treatment is $D = D_Z$. By construction, the potential treatments satisfy Assumption \ref{as:default} with default choice $\boldsymbol j^{\ast} = D_{z_0}$, and the conditional choice probabilities are $P \{D = d | Z = z_0\} = \frac{1}{J}$ and $P \{D = d | Z = j\} = \frac{3}{4J} + \frac{1}{4} I \{d = j\}$ otherwise. Our outcome variable is determined by a variable $V$. We set $V \mid (D = d, Z = z) \sim N(1, 4)$ except when $d = 0$ and $z \in \{1, \dots, J - 1\}$. For $1 \leq j \leq J - 1$, we set $V \mid (D = 0, Z = j) \sim (1 - \theta) N(1, 4) + \theta N(1, 1/4)$ for $0 \leq \theta \leq 1$, so that as $\theta$ increases, the conditional distribution is more concentrated around 1. The choice probabilities, however, do not vary with $\theta$.
 
{
\spacingset{\TABLESPACING}
\begin{table}[ht!]
    \centering\normalsize
    \begin{tabular}{rcccccc@{\hspace{8pt}}cccccc}
    \toprule
    \multicolumn{13}{c}{(a) $\mathsf{base} \notin \mathcal Z$, $J = 3$} \\
    \addlinespace[4pt]
     & \multicolumn{6}{c}{\textbf{Closed Form}} & \multicolumn{6}{c}{\textbf{Latent}} \\
    \cmidrule(lr){2-7} \cmidrule(lr){8-13}
     &  & \multicolumn{2}{c}{$H_0$} & \multicolumn{3}{c}{$H_1$} &   & \multicolumn{2}{c}{$H_0$} & \multicolumn{3}{c}{$H_1$} \\
    \cmidrule(lr){3-4} \cmidrule(lr){5-7} \cmidrule(lr){9-10} \cmidrule(lr){11-13}
    $n$ &  & $\theta = 0$ & $c_\ast$ & $0.35$ & $0.50$ & $0.75$ &  & $0$ & $c_\ast$ & $0.35$ & $0.50$ & $0.75$ \\
    \cmidrule(lr){1-13}
    500 &  & 0.4 & 3.0 & 8.6 & 25.1 & 60.3 &  & 0.0 & 1.7 & 4.3 & 14.0 & 44.3 \\
    1000 &  & 0.0 & 3.1 & 14.4 & 38.8 & 88.7 &  & 0.0 & 1.8 & 6.5 & 27.6 & 78.0 \\
    2000 &  & 0.0 & 4.0 & 25.0 & 71.3 & 99.5 &  & 0.0 & 2.7 & 17.4 & 59.2 & 99.2 \\
    \addlinespace[12pt]
    \multicolumn{13}{c}{(b) $\mathsf{base} \in \mathcal Z$, $J = 4$} \\
    \addlinespace[4pt]
     & \multicolumn{6}{c}{\textbf{Closed Form}} & \multicolumn{6}{c}{\textbf{Latent}} \\
    \cmidrule(lr){2-7} \cmidrule(lr){8-13}
     &  & \multicolumn{2}{c}{$H_0$} & \multicolumn{3}{c}{$H_1$} &   & \multicolumn{2}{c}{$H_0$} & \multicolumn{3}{c}{$H_1$} \\
    \cmidrule(lr){3-4} \cmidrule(lr){5-7} \cmidrule(lr){9-10} \cmidrule(lr){11-13}
    $n$ &  & $\theta = 0$ & $c_\ast$ & $0.50$ & $0.75$ & $1$ &  & $0$ & $c_\ast$ & $0.50$ & $0.75$ & $1$ \\
    \cmidrule(lr){1-13}
    500 &  & 0.4 & 1.1 & 6.1 & 19.8 & 41.3 &  & 0.2 & 0.7 & 2.8 & 10.5 & 28.3 \\
    1000 &  & 0.2 & 1.7 & 14.4 & 45.2 & 78.8 &  & 0.1 & 0.4 & 6.6 & 29.6 & 65.8 \\
    2000 &  & 0.1 & 1.7 & 31.5 & 81.3 & 99.5 &  & 0.0 & 0.5 & 18.2 & 67.3 & 96.4 \\
    \bottomrule
    \end{tabular}
    \caption{rejection rates (\%) for $Y = I \{0 < V\leq2\}$ without and with a base state.}
    \label{table:sims-finite}
\end{table}
}

In this section, we set the observed outcome to $Y = I \{0 < V \leq 2\}$. Let $a = 2 \Phi(1/2) - 1$ and $b = 2 \Phi(2) - 1$, where $\Phi$ is the standard normal distribution function. When $\mathsf{base} \notin \mathcal Z$, the sharp inequalities are \eqref{eq:ineq-Y} and \eqref{eq:encourage-Y}; when $\mathsf{base} \in \mathcal Z$, they are \eqref{eq:encourage-Y} and \eqref{eq:0better-Y}. As verified in Appendix D.1, all of the inequalities hold for $Y$ if and only if $\theta \leq c_\ast := \frac{a}{3(b - a)}$. Table \ref{table:sims-finite} reports the rejection rates for both specifications. In each panel, we consider $\theta = 0$ in the interior of the null, $\theta = c_\ast$ on its boundary, and three values of $\theta$ in the alternative.\footnote{We omit the results not using the outcome because the choice probabilities satisfy the restrictions in \eqref{eq:ineq} or \eqref{eq:0better} for every $\theta \in [0, 1]$; in fact, they do not change with $\theta$.} At all sample sizes, both Closed Form and Latent control size, and Closed Form has higher power.

We then consider the setting with one-sided noncompliance in Theorem \ref{thm:onesided}, with $\mathcal Z = \{\mathsf{base}, 1, \dots, J - 1\}$ and $J \in \{3, 4\}$. We use the same construction as above, except that we set $D_{\mathsf{base}} \equiv 0$. As we verify in Appendix D.1, the inequalities \eqref{eq:onesided-testable} also hold if and only if $\theta \leq c_\ast$. Table S.2 in the supplement reports the rejection rates. In this design, both Closed Form and Latent overreject on the boundary $\theta = c_\ast$, although Latent does so less severely. The higher power of Closed Form versus Latent in this simulation setting should therefore be interpreted with caution.\footnote{The issue likely arises because some entries of the ``restricted estimator'' in \cite{fang2023inference} are strictly positive when the corresponding inequalities are actually binding in the population. In the one-sided noncompliance setting, half of the Closed Form inequalities bind at $\theta = c_\ast$. As a diagnostic, for $J = 3$ and $n = 1000$, if we set the entries of the restricted estimators in \cite{fang2023inference} corresponding to the binding population inequalities to zero, then the rejection rate for Closed Form drops to around 5.0\%.}



\subsection{Continuous Outcomes} \label{sec:sims-continuous}
We now turn to continuous outcomes. We use the same specification as in Section \ref{sec:sims-finite} with $\mathsf{base} \in \mathcal Z$ and $J = 4$ but take the observed outcome to be $Y = V$. As we verify in Appendix D.2, the inequalities \eqref{eq:encourage-Y} and \eqref{eq:0better-Y} hold for every Borel set $B \subseteq \mathbf R$ if and only if $\theta \leq 1/9$. Following Lemma \ref{lem:cells}, we test the inequalities for two partitions of the outcome space: $\mathcal I_1 = \{(-\infty, 1], (1, \infty)\}$ and its refinement $\mathcal I_2 = \{(-\infty, 0], (0, 1], (1, 2], (2, \infty)\}$. As we verify in Appendix D.2, the cell inequalities in Lemma \ref{lem:cells} for $\mathcal I_1$ hold for every $\theta \in [0, 1]$, whereas those for $\mathcal I_2$ hold if and only if $\theta \leq c_\ast$.

{
\spacingset{\TABLESPACING}
\begin{table}[ht!]
    \centering\normalsize
    \begin{tabular}{rlcccccc@{\hspace{8pt}}cccccc}
    \toprule
    \multicolumn{14}{c}{$\mathsf{base} \in \mathcal Z$, $J = 4$} \\
    \addlinespace[4pt]
     &  & \multicolumn{6}{c}{\textbf{Closed Form}} & \multicolumn{6}{c}{\textbf{Latent}} \\
    \cmidrule(lr){3-8} \cmidrule(lr){9-14}
     &  &  & \multicolumn{2}{c}{$H_0$} & \multicolumn{3}{c}{$H_1$} &   & \multicolumn{2}{c}{$H_0$} & \multicolumn{3}{c}{$H_1$} \\
    \cmidrule(lr){4-5} \cmidrule(lr){6-8} \cmidrule(lr){10-11} \cmidrule(lr){12-14}
    $n$ &  &  & $\theta = 0$ & $1/9$ & $0.50$ & $0.75$ & $1$ &  & $0$ & $1/9$ & $0.50$ & $0.75$ & $1$ \\
    \cmidrule(lr){1-14}
    500 & No outcome &  & 0.0 & 0.0 & 0.0 & 0.0 & 0.0 &  & 0.2 & 0.2 & 0.2 & 0.2 & 0.2 \\
     & Two bins &  & 0.2 & 0.2 & 0.2 & 0.2 & 0.2 &  & 0.1 & 0.1 & 0.1 & 0.1 & 0.1 \\
     & Four bins &  & 1.0 & 1.1 & 4.3 & 11.9 & 28.3 &  & 0.4 & 0.3 & 1.8 & 4.3 & 11.7 \\
    \addlinespace[6pt]
    1000 & No outcome &  & 0.1 & 0.1 & 0.1 & 0.1 & 0.1 &  & 0.1 & 0.1 & 0.1 & 0.1 & 0.1 \\
     & Two bins &  & 0.1 & 0.1 & 0.1 & 0.1 & 0.1 &  & 0.0 & 0.0 & 0.0 & 0.0 & 0.0 \\
     & Four bins &  & 0.0 & 0.0 & 6.2 & 23.5 & 58.8 &  & 0.0 & 0.0 & 2.4 & 10.6 & 33.7 \\
    \addlinespace[6pt]
    2000 & No outcome &  & 0.0 & 0.0 & 0.0 & 0.0 & 0.0 &  & 0.0 & 0.0 & 0.0 & 0.0 & 0.0 \\
     & Two bins &  & 0.0 & 0.0 & 0.0 & 0.0 & 0.0 &  & 0.0 & 0.0 & 0.0 & 0.0 & 0.0 \\
     & Four bins &  & 0.0 & 0.4 & 17.3 & 58.5 & 93.3 &  & 0.0 & 0.0 & 6.7 & 38.3 & 78.9 \\
    \bottomrule
    \end{tabular}
    \caption{rejection rates (\%) for continuous outcomes, $\mathsf{base} \in \mathcal Z$, $J = 4$. Hypotheses concern the full distribution of $Y$.}
    \label{table:sims-continuous}
\end{table}
}

Table \ref{table:sims-continuous} compares tests without an outcome with tests using each partition. The hypotheses are labeled according to whether the full distribution of $(Y, D, Z)$ belongs to the null hypothesis in \eqref{eq:null}, so $\theta = 1/9$ is the new boundary. The four-bin inequalities remain strictly slack at $\theta = 1/9$. The conditional probabilities without an outcome and with two bins do not change with $\theta$, so these tests cannot distinguish the alternatives from the null.\footnote{Moreover, under the common simulation draws, the results of the test are identical across $\theta$, thus leading to identical rejection rates.} The four-bin inequalities are violated at all three alternatives, and the test based on them has nontrivial power that increases with $\theta$. In Table S.3 in the appendix, we further consider eight bins with 100 replications. The rejection rates are higher than with four bins for both approaches, and the difference in power between Closed Form and Latent is larger at $\theta = 0.50$ and $0.75$.

Finally, Table S.4 in the appendix reports the matrix dimensions and runtime. The runtime for both approaches is similar for low-dimensional examples, with Latent often being slightly faster, but Closed Form is noticeably faster for continuous outcomes with a larger number of discretization bins. At $n = 2000$, Closed Form is about twice as fast as Latent with four bins and about 20 times as fast with eight bins.

\section{Empirical Application} \label{sec:empirical}
We apply our methods to the dataset from \cite{behaghel2013robustness,behaghel2014private}. They study an RCT with three job search counseling programs in France. In this section, we use $\mathsf a, \mathsf b, \mathsf c$ instead of $0, 1, 2$ to denote these three programs as well as the corresponding instrument values. In their setting, $Z = \mathsf a$ denotes assignment to the usual public program without intensive counseling, which serves as the control group (note that in what follows, we consider both the cases where we interpret $Z = \mathsf a$ as a base state and as an encouraged state); $Z = \mathsf b$ denotes assignment to the public program with intensive counseling; and $Z = \mathsf c$ denotes assignment to the private program with intensive counseling. $D \in \{\mathsf a, \mathsf b, \mathsf c\}$ denotes participation in the corresponding programs. Assumption \ref{as:exog} holds because $Z$ is randomly assigned. As Table \ref{table:behaghel-dz} shows, some job seekers assigned to each program entered one of the other two programs. Assignment is therefore an encouragement that one may or may not take up.

{
\spacingset{\TABLESPACING}
\begin{table}[ht!]
    \centering
    \begin{tabular}{cccc}
    \toprule
        $Z \backslash D$ & $\mathsf a$ & $\mathsf b$ & $\mathsf c$ \\
       \midrule
       $\mathsf a$ & 0.9620 & 0.0036 & 0.0344 \\
       $\mathsf b$ & 0.6592 & 0.3170 & 0.0238 \\
       $\mathsf c$ & 0.5399 & 0.0044 & 0.4557 \\
       \bottomrule
    \end{tabular}
    \caption{Estimated values of $P \{D = d | Z = z\}$.}
    \label{table:behaghel-dz}
\end{table}
}

We consider three outcomes in their dataset: ``EMPLOI 6MOIS,'' which indicates exit from PES registers to employment; ``EMPLOI AR110 6MOIS,'' which indicates any employment; and ``SUCCES OPP 6MOIS,'' which indicates employment eligible for payment. All three outcomes are binary and were measured six months after the beginning of the programs. To recover a causal interpretation of the IV estimand, \cite{behaghel2013robustness} impose an assumption called ``extended monotonicity,'' which we show in Appendix B to be strictly stronger than interpreting $Z = \mathsf a$ as a base state in our model. As a result, we focus on testing \eqref{eq:null} in two specifications of the model: the base-state specification, in which $Z = \mathsf a$ is interpreted as a base state and $Z = \mathsf b, \mathsf c$ encourage programs $\mathsf b$ and $\mathsf c$, and the no-base-state specification, in which $Z = \mathsf a, \mathsf b, \mathsf c$ encourage programs $\mathsf a$, $\mathsf b$, and $\mathsf c$.
The base-state test indicates whether assignment to the control group can be assumed not to encourage participation in program $\mathsf a$. Because all outcomes have finite supports, we test the inequalities following Lemma \ref{lem:cells} applied to singletons of $\mathcal Y$. We consider each outcome separately and also jointly as a vector-valued outcome. For comparison, we also test \eqref{eq:ineq} or \eqref{eq:0better} without outcomes. All tests apply \cite{fang2023inference} with $5{,}000$ bootstrap draws and the tuning parameter in Section \ref{sec:sims}. Table \ref{table:behaghel} reports the $p$-values in percentages. Without outcomes, neither specification is rejected at the 10\% level. With each outcome separately or jointly, the base-state specification is rejected at the 5\% level, but the no-base-state specification is not rejected at the 10\% level.

{
\spacingset{\TABLESPACING}
\begin{table}[ht!]
    \centering
    \begin{tabular}{ccc}
    \toprule
    $Y$ & no-base-state & base state \\
    \midrule
    - & 100.00 & 23.34 \\
    EMPLOI 6MOIS & 100.00 & 0.60 \\
    EMPLOI AR110 6MOIS & 100.00 & 0.00 \\
    SUCCES OPP 6MOIS & 100.00 & 1.62 \\
    Joint outcome vector & 100.00 & 0.02 \\
    \bottomrule
    \end{tabular}
    \caption{$p$-values in percentages for testing \eqref{eq:ineq} or \eqref{eq:0better} without $Y$, and the finite-outcome restrictions in Lemma \ref{lem:cells} with $Y$, using \cite{fang2023inference}.}
    \label{table:behaghel}
\end{table}
}

To detect the substitution patterns responsible for rejection, we first calculate the point estimates of all moments in the tests. Among them, moments of the form $P \{Y = y, D = \mathsf b | Z = \mathsf c\} \leq P \{Y = y, D = \mathsf b | Z = \mathsf a\}$ are most consistently violated across the outcomes. Next, we apply the stepdown multiple hypothesis testing procedure in \cite{romano2018multiple} to formally detect which moments are violated. We coarsen the joint outcome vector into the binary indicator $U$, defined as the maximum of the three outcomes, which indicates employment according to at least one of the three definitions. We test all inequalities in \eqref{eq:encourage-Y} and \eqref{eq:0better-Y} for the outcome $U$ as well as the inequalities without the outcome in \eqref{eq:0better}. The procedure rejects the base-state specification with a $p$-value of $0.0278$, and the rejection comes from $P \{U = 1, D = \mathsf b | Z = \mathsf c\} \leq P \{U = 1,D = \mathsf b | Z = \mathsf a\}$. To understand the violated moment, note that in Assumption \ref{as:default}, $D_{\mathsf c} = \mathsf b$ and $D_{\mathsf c} \in \{D_{\mathsf a},\mathsf c\}$ imply $D_{\mathsf a} = \mathsf b$. The rejection therefore points to the substitution pattern $D_{\mathsf c} = \mathsf b$ and $D_{\mathsf a} \neq \mathsf b$, choosing program $\mathsf b$ when encouraged towards program $\mathsf c$ but not choosing program $\mathsf b$ when assigned to program $\mathsf a$. Under the no-base-state specification, changing $Z = \mathsf a$ to $Z = \mathsf c$ increases the appeal of program $\mathsf c$ and decreases that of program $\mathsf a$, allowing a switch to program $\mathsf b$. In contrast, under the base-state specification, it only increases the appeal of program $\mathsf c$, so any switch must be to $\mathsf c$. However, we note that the relevant cells contain 55 observations under $Z = \mathsf c$ but only two under $Z = \mathsf a$, so the results should be interpreted cautiously.

Maintaining the no-base-state specification, Theorem \ref{thm:subs} shows that the sharp lower bound on the probability of the substitution pattern detected above, $\{D_{\mathsf c} = \mathsf b,D_{\mathsf a} \neq \mathsf b\}$, is
\[ \sum_{y \in \mathcal Y} \big ( P \{Y = y, D = \mathsf b | Z = \mathsf c\} - P \{Y = y, D = \mathsf b | Z = \mathsf a\} \big )_+~. \]
Using the joint outcome vector, the estimated lower bound is $0.1309\%$, and the lower end of the one-sided 95\% confidence interval inverting the test in \cite{fang2023inference} is $0.073\%$. The same sparse cells contribute to this bound, so the result should be interpreted cautiously.

\clearpage
\appendix
\setcounter{equation}{0}
\setcounter{table}{0}
\renewcommand{\theequation}{S.\arabic{equation}}
\renewcommand{\thetable}{S.\arabic{table}}
\renewcommand{\theHequation}{S.\arabic{equation}}
\renewcommand{\theHtable}{S.\arabic{table}}

\section{Proofs of Main Results}

\subsection{Proof of Theorem \ref{thm:sharp}(a)}
To establish the inequalities, fix $z(j) \in \mathcal Z(j)$ for $0 \leq j \leq J - 1$. For $0 \leq j \leq J - 1$, $D_{z(j)} = j$ implies $\boldsymbol{j}^{\ast} = j$. If $z(j) \neq \mathsf{base}$, then this follows from \eqref{eq:default} and $z(j) \neq j$; if $z(j) = \mathsf{base}$, then this follows from \eqref{eq:default-base}. Therefore,
\[ \{D_{z(0)} = 0\}, \dots, \{D_{z(J - 1)} = J - 1\} \]
are disjoint events, so
\[ \sum_{0 \leq j \leq J - 1} Q \{D_{z(j)} = j\} \leq 1~. \]
The desired conclusion now follows from Assumption \ref{as:exog}. \qed

\subsection{Proof of Theorem \ref{thm:sharp}(b)}
We construct $Q^\ast$ that satisfies Assumptions \ref{as:exog} and \ref{as:default}, for which $P = Q^\ast T^{-1}$. Recall $\mathcal Z(j) = \mathcal Z \setminus \{j\}$. Because $|\mathcal Z| \geq 2$, $\mathcal Z(j)$ is nonempty for all $0 \leq j \leq J - 1$. In what follows, it is often useful to rewrite the requirement that \eqref{eq:ineq} holds for every collection of $z(j) \in \mathcal Z(j)$, $0 \leq j \leq J - 1$ as
\begin{equation} \label{eq:ineq-compact}
    \sum_{0 \leq j \leq J - 1} \max_{z \in \mathcal Z(j)} P \{D = j | Z = z\} \leq 1~.
\end{equation}
If $\mathsf{base} \in \mathcal Z$, then $\mathsf{base} \in \mathcal Z(j)$ for all $0 \leq j \leq J - 1$. We have by definition that
\[ \sum_{0 \leq j \leq J - 1} P \{D = j | Z = \mathsf{base}\} = 1~. \]
Together with \eqref{eq:ineq-compact} and probabilities being nonnegative, it implies that for all $0 \leq j \leq J - 1$,
\begin{equation} \label{eq:base-max}
P \{D = j | Z = \mathsf{base}\} = \max_{z \in \mathcal Z(j)} P \{D = j | Z = z\}~.    
\end{equation}

The construction of $Q^\ast$ is done in $J$ blocks indexed by $0 \leq j \leq J - 1$ and a final diagonal block. For block $j$, where $0 \leq j \leq J - 1$, let $\{z_1(j), \dots, z_{|\mathcal Z(j)|}(j)\} = \mathcal Z(j)$ be such that
\[ P \{D = j | Z = z_1(j)\} \leq \dots \leq P \{D = j | Z = z_{|\mathcal Z(j)|}(j)\}~. \]
If $\mathsf{base} \in \mathcal Z$, then the previous discussion shows that we can always define $z_{|\mathcal Z(j)|}(j) = \mathsf{base}$, which we do including in the case of ties. We break the remaining ties according to the numerical values of $z$ from the smallest to the largest. Our construction for this fixed $j$ consists of the following steps:

Step 1: define
\begin{equation} \label{eq:step-1}
Q^\ast \{D_z = j \text{ for all } z \in \mathcal Z\} = P \{D = j | Z = z_1(j)\}~.
\end{equation}
    
Step $i$ for $2 \leq i \leq |\mathcal Z(j)|$: define
\begin{multline} \label{eq:step-i}
Q^\ast \big \{D_{z_1(j)} = z_1(j), \dots, D_{z_{i - 1}(j)} = z_{i - 1}(j), D_z = j \text{ for } z \in \mathcal Z \setminus \{z_1(j), \dots, z_{i - 1}(j)\} \big \} \\
= P \{D = j | Z = z_i(j)\} - P \{D = j | Z = z_{i - 1}(j)\}~.
\end{multline}
Note that the labels in \eqref{eq:step-i} are admissible even if $\mathsf{base} \in \mathcal Z$, because in that case $z_{|\mathcal Z(j)|}(j) = \mathsf{base}$ and in \eqref{eq:step-i} the treatment never takes $z_{|\mathcal Z(j)|}(j)$ as its value.

Carry out these steps for blocks indexed by $0 \leq j \leq J - 1$. The final block is constructed as follows. If $\mathsf{base} \notin \mathcal Z$, then we define the ``diagonal'' as
\begin{equation} \label{eq:diagonal}
Q^\ast \{D_z = z \text{ for } z \in \mathcal Z\} = 1 - \sum_{0 \leq j \leq J - 1} \max_{z \in \mathcal Z(j)} P \{D = j | Z = z\}~,
\end{equation}
where the right-hand side is nonnegative because of \eqref{eq:ineq-compact}. If $\mathsf{base} \in \mathcal Z$, then the right-hand side is zero because of \eqref{eq:base-max}, so this step of the construction is vacuous. For all other vectors $(d_z: z \in \mathcal Z)$, define $Q^\ast \{D_z = d_z \text{ for all } z \in \mathcal Z\} = 0$.

The idea of the construction is that \eqref{eq:step-1}--\eqref{eq:step-i} completely specify the probability of potential treatments where $j$ is chosen for at least one instrument value in $\mathcal Z(j)$; when $\mathsf{base} \notin \mathcal Z$, \eqref{eq:diagonal} closes the gap by specifying the probability that one complies with all values of the instrument. 

Finally, for each $z$ and $(d_k: k \in \mathcal Z)$, define
\begin{equation} \label{eq:contruct-exog}
Q^\ast \{D_k = d_k \text{ for all } k \in \mathcal Z, Z = z\} = Q^\ast \{D_k = d_k \text{ for all } k \in \mathcal Z\} P \{Z = z\}~.
\end{equation}

We now verify that $Q^\ast$ is a probability measure. All assigned probabilities are nonnegative by construction. For each $j$, clearly the events in \eqref{eq:step-1}--\eqref{eq:step-i} are disjoint. In addition, they are disjoint across blocks $0 \leq j \leq J - 1$: on the events in block $j$, $j$ is chosen for at least one instrument value in $\mathcal Z(j)$, which cannot happen in another block. In another block $j' \neq j$, choosing $j$ is possible only when $z = j$, which is excluded from $\mathcal Z(j)$. None of these events is the diagonal because on the diagonal, every choice equals its instrument value, so choosing $j$ under an assignment in $\mathcal Z(j)$ is likewise impossible. For $0 \leq j \leq J - 1$, the sum of the right-hand sides of \eqref{eq:step-1}--\eqref{eq:step-i} is
\[ P \{D = j | Z = z_{|\mathcal Z(j)|}(j)\} = \max_{z \in \mathcal Z(j)} P \{D = j | Z = z\}~. \]
Therefore, across $0 \leq j \leq J - 1$, the sum of the right-hand sides of \eqref{eq:step-1}--\eqref{eq:diagonal} is 1. As a result, $Q^\ast$ is a probability measure.

Next, we verify that $Q^\ast$ satisfies Assumptions \ref{as:exog} and \ref{as:default}. First note Assumption \ref{as:exog} holds by \eqref{eq:contruct-exog}. By construction, in the event in \eqref{eq:step-1} and \eqref{eq:step-i}, $j$ is chosen for at least one value of $z \in \mathcal Z(j)$, so on this event we can define the default value $\boldsymbol{j}^{\ast} = j$; if $\mathsf{base} \in \mathcal Z$, then by construction $D_{\mathsf{base}} = j$. On the diagonal event in \eqref{eq:diagonal}, we can define $\boldsymbol{j}^{\ast}$ as any choice, and for concreteness we define $\boldsymbol{j}^{\ast} = 0$. For $\boldsymbol{j}^{\ast}$ defined this way, $Q^\ast$ satisfies \eqref{eq:default}, and if $\mathsf{base} \in \mathcal Z$, then $Q^\ast$ also satisfies \eqref{eq:default-base}, so in summary $Q^\ast$ satisfies Assumption \ref{as:default}.

To conclude the proof, we verify that $P = Q^\ast T^{-1}$, i.e., \eqref{eq:P-Q} holds. Fix $0 \leq j \leq J - 1$ and $z \in \mathcal Z(j)$, and note that $z = z_i(j)$ for some $1 \leq i \leq |\mathcal Z(j)|$. Under $Q^\ast$, the only events with nonzero probability on which $D_z = j$ for $z \in \mathcal Z(j)$ are in block $j$ of the construction. If $i \geq 2$, then
\begin{align*}
& Q^\ast \{D_{z_i(j)} = j\} \\
& = Q^\ast \{D_k = j \text{ for } k \in \mathcal Z\} + \dots \\
& \hspace{1.5em} + Q^\ast \big \{D_{z_1(j)} = z_1(j), \dots, D_{z_{i - 1}(j)} = z_{i - 1}(j), D_z = j \text{ for all } z \in \mathcal Z \setminus \{z_1(j), \dots, z_{i - 1}(j)\} \big \} \\
& = P \{D = j | Z = z_1(j)\} + P \{D = j | Z = z_2(j)\} - P \{D = j | Z = z_1(j)\} + \dots \\
& \hspace{1.5em} + P \{D = j | Z = z_i(j)\} - P \{D = j | Z = z_{i - 1}(j)\} \\
& = P \{D = j | Z = z_i(j)\}~,
\end{align*}
while for $i = 1$ the equality holds immediately by \eqref{eq:step-1}. Therefore, we have verified $Q^\ast \{D_z = j\} = P \{D = j | Z = z\}$ for all $0 \leq j \leq J - 1$ and $z \in \mathcal Z(j)$. Because $\mathcal Z(j) = \mathcal Z$ for $j < J_0$, it remains to verify $Q^\ast \{D_k = k\} = P \{D = k | Z = k\}$ for $J_0 \leq k \leq J - 1$. To do so, note by the equalities that we have verified above,
\begin{align}
\nonumber Q^\ast \{D_k \neq k\} & = \sum_{0 \leq j \leq J - 1: j \neq k} Q^\ast \{D_k = j\} \\
\label{eq:knotk} & = \sum_{0 \leq j \leq J - 1: j \neq k} P \{D = j | Z = k\} = 1 - P \{D = k | Z = k\}~.   
\end{align} 
Because $Q^\ast$ is a probability measure, $Q^\ast \{D_k = k\} = 1 - Q^\ast \{D_k \neq k\} = P \{D = k | Z = k\}$, and \eqref{eq:P-Q} is satisfied. Because the marginal distributions of $Z$ are the same under $Q^\ast$ and $P$, $P = Q^\ast T^{-1}$. \qed

\subsection{Proof of Corollary \ref{cor:0better}}
First note that \eqref{eq:0better} holds with equality when $k = \mathsf{base}$. For $0 \leq j \leq J - 1$ and $k \in \mathcal Z(j) \setminus \{\mathsf{base}\}$, set $z(j) = k$ and $z(\ell) = \mathsf{base} \in \mathcal Z(\ell)$ for all $\ell \in \{0, \dots, J - 1\} \setminus \{j\}$ in \eqref{eq:ineq}, and we get
\[ \sum_{\ell \neq j} P \{D = \ell | Z = \mathsf{base}\} + P \{D = j | Z = k\} \leq 1~, \]
which implies \eqref{eq:0better}. Conversely, suppose \eqref{eq:0better} holds for all $0 \leq j \leq J - 1$ and $k \in \mathcal Z(j)$. For every collection $z(j) \in \mathcal Z(j)$ for $0 \leq j \leq J - 1$, \eqref{eq:0better} implies
\[ \sum_{0 \leq j \leq J - 1} P \{D = j | Z = z(j)\} \leq \sum_{0 \leq j \leq J - 1} P \{D = j | Z = \mathsf{base}\} = 1~, \]
so \eqref{eq:ineq} follows. \qed

\subsection{Proof of Theorem \ref{thm:sharp-Y}}
We first prove the necessity of parts (a) and (b). For each $0 \leq j \leq J - 1$, let $\{B_z(j): z \in \mathcal Z(j)\}$ be Borel sets that form a partition of $\mathcal Y$. If $D_z = j$ for $z \in \mathcal Z(j)$, then $\boldsymbol{j}^{\ast} = j$ by Assumption \ref{as:default}. Therefore,
\[ (\{Y_j \in B_z(j), D_z = j\}: 0 \leq j \leq J - 1, z \in \mathcal Z(j)) \]
are disjoint because different values of $j$ imply different values of $\boldsymbol{j}^{\ast}$ and for the same $j$, $B_z(j)$ are disjoint across $z \in \mathcal Z(j)$. Therefore, the sum of their probabilities is at most one, and Assumption \ref{as:exog-Y} implies the inequalities in \eqref{eq:ineq-Y}. For a Borel set $B \subseteq \mathcal Y$, $J_0 \leq j \leq J - 1$, and $k \in \mathcal Z(j)$, it follows from Assumption \ref{as:default} that $D_k = j$ implies $D_j = j$. Therefore,
\[ \{Y_j \in B, D_k = j\} \implies \{Y_j \in B, D_j = j\}~, \]
and
\eqref{eq:encourage-Y} follows.

We now prove the sufficiency of parts (a) and (b). We construct $Q^\ast$ that satisfies Assumptions \ref{as:exog-Y} and \ref{as:default} and such that $P = Q^\ast T^{-1}$. To begin, define the common dominating measure $\nu$ by
\[ \nu(B) = \frac{1}{|\mathcal Z|} \sum_{z \in \mathcal Z} P \{Y \in B | Z = z\} \]
for all Borel sets $B \subseteq \mathcal Y$. Let $p_j(\cdot | z)$ denote a Borel-measurable version of the Radon-Nikodym derivative of $P \{Y \in \cdot, D = j | Z = z\}$ with respect to $\nu$. The densities are nonnegative and finite $\nu$-almost surely. Moreover, \eqref{eq:encourage-Y} implies $p_j(y | k) \leq p_j(y | j)$ $\nu$-almost surely for $J_0 \leq j \leq J - 1$ and $k \in \mathcal Z(j)$. Therefore, by setting all densities equal to zero on a common $\nu$-null Borel set, we assume that they are nonnegative and finite for every $y$ and the inequalities hold for all $y \in \mathcal Y$. 

For each $0 \leq j \leq J - 1$ and each $y \in \mathcal Y$, let $\{z_1(y, j), \dots, z_{|\mathcal Z(j)|}(y, j)\} = \mathcal Z(j)$ be such that
\[ p_j(y | z_1(y, j)) \leq \dots \leq p_j(y | z_{|\mathcal Z(j)|}(y, j))~. \]
If there are ties, we break them according to the ordering of $z$ from the smallest to the largest, and if the ties involve $\mathsf{base}$, we put it last. Each function $z_i(\cdot, j)$ is Borel measurable because it is determined by finitely many comparisons of Borel-measurable functions. In \eqref{eq:ineq-Y}, taking $B_z(j) = \{y \in \mathcal Y: z = z_{|\mathcal Z(j)|}(y, j)\}$, we get
\begin{equation} \label{eq:ineq-Y-compact}
\sum_{0 \leq j \leq J - 1} \int_{\mathcal Y} \max_{z \in \mathcal Z(j)} p_j(y | z) d \nu(y) \leq 1~. 
\end{equation}
If $\mathsf{base} \in \mathcal Z$, then $\mathsf{base} \in \mathcal Z(j)$ for all $0 \leq j \leq J - 1$. Because all densities are nonnegative and
\[ \sum_{0 \leq j \leq J - 1} \int_{\mathcal Y} p_j(y | \mathsf{base}) d \nu(y) = \sum_{0 \leq j \leq J - 1} P \{D = j | Z = \mathsf{base}\} = 1~, \]
we see from \eqref{eq:ineq-Y-compact} that for all $0 \leq j \leq J - 1$, $\nu$-almost surely,
\[ p_j(y | \mathsf{base}) = \max_{z \in \mathcal Z(j)} p_j(y | z)~. \]
Set all densities equal to zero on one common $\nu$-null Borel set where any of these equalities fails, and redefine the ordering there using the same tie-breaking rule. We then have $z_{|\mathcal Z(j)|}(y, j) = \mathsf{base}$ for every $y \in \mathcal Y$ and $0 \leq j \leq J - 1$.

As in the proof of Theorem \ref{thm:sharp}(b), the construction is done in $J$ blocks indexed by $0 \leq j \leq J - 1$ and a final diagonal block. For $j < J_0$, define $\lambda_j(y) \equiv 1$. For $J_0 \leq j \leq J - 1$, define $\lambda_j(y) \equiv 1$ (or any measurable and nonnegative function such that $\int_{\mathcal Y} \lambda_j(y) d \nu(y) = 1$) if $\int_{\mathcal Y} \big ( p_j(y | j) - \max_{z \in \mathcal Z(j)} p_j(y | z) \big ) d \nu(y) = 0$, and otherwise define
\begin{equation} \label{eq:lambda}
\lambda_j(y) = \frac{\displaystyle p_j(y | j) - \max_{z \in \mathcal Z(j)} p_j(y | z)}{\displaystyle \int_{\mathcal Y} \big (p _j(y | j) - \max_{z \in \mathcal Z(j)} p_j(y | z) \big ) d \nu(y)}~.   
\end{equation}
Note that $\lambda_j \geq 0$ for all $j$, where for $J_0 \leq j \leq J - 1$ it follows from \eqref{eq:encourage-Y}. By construction, $\int_{\mathcal Y} \lambda_j(y) d \nu(y) = 1$.

For each $(d_z: z \in \mathcal Z) \in \{0, \dots, J - 1\}^{\mathcal Z}$ and Borel sets $B_0, \dots, B_{J - 1} \subseteq \mathcal Y$, we write
\[ Q_{(d_z: z \in \mathcal Z)}^\ast(B_0 \times \dots \times B_{J - 1}) = Q^\ast \{Y_0 \in B_0, \dots, Y_{J - 1} \in B_{J - 1}, D_z = d_z \text{ for all } z \in \mathcal Z\}~. \]
Let $\nu^J = \nu \times \dots \times \nu$ denote the product measure on $\mathcal Y^J$ with each marginal equal to $\nu$. We will construct nonnegative Borel-measurable functions $q_{(d_z: z \in \mathcal Z)}^\ast$ and define for Borel sets $B_0, \dots, B_{J - 1} \subseteq \mathcal Y$ that
\[ Q_{(d_z: z \in \mathcal Z)}^\ast(B_0 \times \dots \times B_{J - 1}) = \int_{B_0}\dots\int_{B_{J - 1}} q_{(d_z: z \in \mathcal Z)}^\ast(y_0, \dots, y_{J - 1}) d \nu(y_0) \dots d \nu(y_{J - 1})~. \]
Finally, as in the proof of Theorem \ref{thm:sharp}(b), we will define the joint distribution $Q^\ast$ by multiplying the marginal distribution of $Z$ under $P$.

Our construction for each $y \in \mathcal Y$ and $0 \leq j \leq J - 1$ consists of the following steps:

Step 1: for each $(y_k \in \mathcal Y: k \neq j)$, define
\[ q_{(j, \dots, j)}^\ast(y_0, \dots, y, \dots, y_{J - 1}) = \prod_{k \neq j} \lambda_k(y_k) p_j(y | z_1(y, j))~. \]
    
Step $i$ for $2 \leq i \leq |\mathcal Z(j)|$: for each $(y_k \in \mathcal Y: k \neq j)$, and for $(d_z: z \in \mathcal Z)$ such that $d_{z_1(y, j)} = z_1(y, j), \dots, d_{z_{i - 1}(y, j)} = z_{i - 1}(y, j)$ and $d_z = j$ for $z \in \mathcal Z \setminus \{z_1(y, j), \dots, z_{i - 1}(y, j)\}$, define
\[ q_{(d_z: z \in \mathcal Z)}^\ast(y_0, \dots, y, \dots, y_{J - 1}) = \prod_{k \neq j} \lambda_k(y_k) \big ( p_j(y | z_i(y, j)) - p_j(y | z_{i - 1}(y, j)) \big )~. \]
The choice $j$ is chosen for at least one instrument value in $\mathcal Z(j)$. Note that the labels are admissible even if $\mathsf{base} \in \mathcal Z$, because in that case $z_{|\mathcal Z(j)|}(y, j) = \mathsf{base}$ and in the steps above, the treatment never takes $z_{|\mathcal Z(j)|}(y, j)$ as its value; moreover, $D_{\mathsf{base}} = j$ on the event above.

Define $\mathcal S_j$ as the set of potential treatment vectors $(d_z: z \in \mathcal Z)$ satisfying $d_z \in \{z, j\}$ for $J_0 \leq z \leq J - 1$ and $d_z = j$ for at least one $z \in \mathcal Z(j)$, together with $d_{\mathsf{base}} = j$ if $\mathsf{base} \in \mathcal Z$. The probabilities are assigned to $\mathcal S_j$ in block $j$ of the construction.

Carry out these steps for blocks indexed by $0 \leq j \leq J - 1$. The final block is constructed as follows. If $\mathsf{base} \notin \mathcal Z$, then we define the ``diagonal'' for each $(y_j \in \mathcal Y: 0 \leq j \leq J - 1)$ by
\[ q_{(J_0, \dots, J - 1)}^\ast(y_0, \dots, y_{J - 1}) = \prod_{0 \leq j \leq J - 1} \lambda_j(y_j) \left ( 1 - \sum_{0 \leq j \leq J - 1} \int_{\mathcal Y} \max_{z \in \mathcal Z(j)} p_j(y | z) d\nu(y) \right )~, \]
which is nonnegative because of \eqref{eq:ineq-Y-compact}. If $\mathsf{base} \in \mathcal Z$, then we have seen that the right-hand side is zero, so this step is vacuous.

Finally, for each $(y_0, \dots,y_{J - 1}) \in \mathcal Y^J$, we set all density values not assigned by the preceding construction equal to zero.

Following similar arguments to those in the proof of Theorem \ref{thm:sharp}(b), it is straightforward to verify $Q^\ast$ is a probability measure and satisfies Assumptions \ref{as:exog-Y} and \ref{as:default}. It remains to verify $P = Q^\ast T^{-1}$. Fix $0 \leq j \leq J - 1$, $z \in \mathcal Z(j)$, and a Borel set $B \subseteq \mathcal Y$. For each $y_j$, there exists $1 \leq i(y_j, j) \leq |\mathcal Z(j)|$ such that $z_{i(y_j, j)}(y_j, j) = z$. Exactly the first $i(y_j, j)$ steps in block $j$ have $d_z = j$, so the sum of the corresponding $q$ is
\[ \sum_{(d_{z'}: z' \in \mathcal Z) \in \mathcal S_j: d_z = j} q^\ast_{(d_{z'}: z' \in \mathcal Z)} (y_0, \dots, y_j, \dots, y_{J -1 }) = \prod_{k \neq j} \lambda_k(y_k) p_{j}(y_j | z)~. \]
Because $\int_{\mathcal Y} \lambda_j(y) d \nu(y) = 1$ for $0 \leq j \leq J - 1$, integrating over $y_j \in B$ and $y_k \in \mathcal Y$ for $k \neq j$, we get
\[ Q^\ast \{Y_j \in B, D_z = j\} = \int_B p_j(y_j | z) d\nu(y_j) = P\{Y \in B, D = j | Z = z\}~. \]

It remains to verify for each Borel set $B \subseteq \mathcal Y$ and $J_0 \leq k \leq J - 1$,
\begin{equation} \label{eq:kk-Y}
Q^\ast \{Y_k \in B, D_k = k\} = P \{Y \in B, D = k | Z = k\}~.    
\end{equation}
To begin with, 
\begin{align*}
    Q^\ast \{Y_k \in B, D_k = k\} = &\underbrace{\sum_{j\neq k} Q^\ast \{Y_k \in B, D_k = k, (D_z: z \in \mathcal Z) \in \mathcal S_j\}}_{\rm (A)} \\
    &+ \underbrace{Q^\ast \{Y_k \in B, D_k = k, (D_z: z \in \mathcal Z) \in \mathcal S_k\}}_{\rm (B)} \\
    &+ \underbrace{Q^\ast \Big \{ Y_k \in B, (D_z: z \in \mathcal Z) \notin \bigcup_{j = 0}^{J - 1} \mathcal S_j \Big \}}_{\rm (C)}~,
\end{align*}
where $\mathcal S_j$ for $0 \leq j \leq J - 1$ was defined in the construction. Here (C) is the contribution of the diagonal $d_j = j$ for $J_0 \leq j \leq J - 1$, which we have shown to equal zero if $\mathsf{base} \in \mathcal Z$.

For (A), fix any $j \neq k$. Because $J_0 \leq k \leq J - 1$, $k \in \mathcal Z(j)$. For any $y_j$, there exists an $i(y_j, j)$ with $z_{i(y_j, j)}(y_j, j) = k$, and exactly the steps $i(y_j, j) < i \leq |\mathcal Z(j)|$ have $d_k = k$. The sum of the corresponding $q$ is
\[ \sum_{(d_z: z \in \mathcal Z) \in \mathcal S_j: d_k = k} q^\ast_{(d_z: z \in \mathcal Z)}(y_0, \dots, y_{J - 1}) = \prod_{j' \neq j} \lambda_{j'}(y_{j'}) \big ( \max_{z \in \mathcal Z(j)} p_j(y_j | z) - p_j(y_j | k) \big )~. \]
Of course, if $k = z_{|\mathcal Z(j)|}(y_j, j)$, then the sum is indeed zero. Integrating over $y_{j'} \in \mathcal Y$ for $j' \neq k$ and $y_k \in B$, we get
\begin{align*}
    &Q^\ast \{Y_k \in B, D_k = k, (D_z: z \in \mathcal Z) \in \mathcal S_j\} \\
    &= \int_{y_k \in B} \lambda_k(y_k) \int_{y_j \in \mathcal Y} \big ( \max_{z \in \mathcal Z(j)} p_j(y_j | z) - p_j(y_j | k) \big ) d\nu(y_j) d\nu(y_k)~.
\end{align*}
Summing over $j \neq k$, (A) equals
\[\int_{y_k \in B} \lambda_k(y_k) \sum_{j \neq k} \bigg( \int_{\mathcal Y} \big ( \max_{z \in \mathcal Z(j)} p_j(y | z) - p_j(y | k) \big ) d\nu(y) \bigg) d\nu(y_k)~. \]
For (B), every vector in $\mathcal S_k$ has $D_k = k$. The sum of the corresponding $q$ is
\[ \sum_{(d_z: z \in \mathcal Z) \in \mathcal S_k} q^\ast_{(d_z: z \in \mathcal Z)}(y_0, \dots, y_k, \dots, y_{J - 1}) = \prod_{j' \neq k} \lambda_{j'}(y_{j'}) \max_{z \in \mathcal Z(k)} p_k(y_k | z)~. \]
Integrating over $y_{j'} \in \mathcal Y$ for $j' \neq k$ and $y_k \in B$, we get
\[ \mathrm{(B)} = \int_{y_k \in B} \max_{z \in \mathcal Z(k)} p_k(y_k | z) d\nu(y_k)~. \]
Similarly,
\[ \mathrm{(C)} = \int_{y_k \in B} \lambda_k(y_k) \left ( 1 -  \sum_{0 \leq j \leq J - 1} \int_{\mathcal Y} \max_{z \in \mathcal Z(j)} p_j(y | z) d \nu(y) \right ) d \nu(y_k)~. \]

Summing (A), (B), and (C), we have
\begin{equation} \label{eq:kk-density}
Q^\ast \{Y_k \in B, D_k = k\} = \int_{y_k \in B} q_k(y_k) d \nu(y_k)~,
\end{equation}
where
\begin{align*}
& q_k(y_k) \\
& = \lambda_k(y_k) \sum_{j \neq k} \int_{\mathcal Y} \big ( \max_{z \in \mathcal Z(j)} p_j(y | z) - p_j(y | k) \big ) d \nu(y) \\
& \hspace{3em} + \lambda_k(y_k) \bigg ( 1 - \sum_{0 \leq j \leq J - 1} \int_{\mathcal Y} \max_{z \in \mathcal Z(j)} p_j(y | z) d \nu(y) \bigg ) \\
& \hspace{3em} + \max_{z \in \mathcal Z(k)} p_k(y_k | z) \\
& = \lambda_k(y_k) \int_{\mathcal Y} \big ( p_k(y | k) - \max_{z \in \mathcal Z(k)} p_k(y | z) \big ) d \nu(y) + \max_{z \in \mathcal Z(k)} p_k(y_k | z) \\
& = p_k(y_k | k)~,
\end{align*}
where the second equality follows because
\[ \int_{\mathcal Y} \Big ( \sum_{j \neq k} p_j(y | k) + p_k(y | k) \Big ) d \nu(y) = 1~, \]
and the last equality holds regardless of whether or not the denominator in the definition of $\lambda_k(y_k)$ is zero. Indeed, if it is not zero, then the equality holds because of the definition of $\lambda_k(y_k)$ in \eqref{eq:lambda}; if it is zero, then $p_k(y_k | k) = \max_{z \in \mathcal Z(k)} p_k(y_k | z)$ for $\nu$-almost every $y_k$, so the equality still holds. The desired result in \eqref{eq:kk-Y} now follows from \eqref{eq:kk-density}, and the proof is complete. \qed

\subsection{Proof of Corollary \ref{cor:0better-Y}}
We need only show that \eqref{eq:ineq-Y} and \eqref{eq:0better-Y} are equivalent when $\mathsf{base} \in \mathcal Z$; the conditions in \eqref{eq:encourage-Y} are retained in both statements. If $z = \mathsf{base}$, then \eqref{eq:0better-Y} holds with equality. Otherwise, fix a Borel set $B \subseteq \mathcal Y$, $0 \leq j \leq J - 1$, and $z \in \mathcal Z(j) \setminus \{\mathsf{base}\}$. In \eqref{eq:ineq-Y}, take $B_z(j) = B$, $B_{\mathsf{base}}(j) = \mathcal Y \setminus B$, and $B_{z'}(j) = \emptyset$ for $z' \in \mathcal Z(j) \setminus \{\mathsf{base}, z\}$. For each $\ell \neq j$, take $B_{\mathsf{base}}(\ell) = \mathcal Y$ and $B_{z'}(\ell) = \emptyset$ for $z' \in \mathcal Z(\ell) \setminus \{\mathsf{base}\}$. Then, we have
\begin{align*}
& P \{Y \in B, D = j | Z = z\} + P \{Y \notin B, D = j | Z = \mathsf{base}\} \\
& \hspace{2in} + \sum_{\ell \neq j} P \{Y \in \mathcal Y, D = \ell | Z = \mathsf{base}\} \leq 1~, 
\end{align*}
from which we obtain \eqref{eq:0better-Y}. Conversely, suppose \eqref{eq:0better-Y} holds for all Borel sets $B \subseteq \mathcal Y$, $0 \leq j \leq J - 1$ and $z \in \mathcal Z(j)$. For each $0 \leq j \leq J - 1$, let $\{B_{z'}(j): z' \in \mathcal Z(j)\}$ be a partition of $\mathcal Y$ consisting of Borel sets. Then,
\begin{align*}
& \sum_{0 \leq j \leq J - 1} \sum_{z' \in \mathcal Z(j)} P \{Y \in B_{z'}(j), D = j | Z = z'\} \\
& \leq \sum_{0 \leq j \leq J - 1} \sum_{z' \in \mathcal Z(j)} P \{Y \in B_{z'}(j), D = j | Z = \mathsf{base}\} = 1~,    
\end{align*}
so \eqref{eq:ineq-Y} holds. \qed

\subsection{Proof of Theorem \ref{thm:subs}}
We first establish the validity of the right-hand side of \eqref{eq:subs} as a lower bound for the left-hand side. Fix $Q \in \mathbf Q^{\mathsf Y}$ such that $P = Q T^{-1}$. Assumption \ref{as:exog-Y} implies that, for every Borel set $C \subseteq B$,
\begin{align*}
& P \{Y \in C, D = j | Z = z\} - P \{Y \in C, D = j | Z = z'\} \\
&= Q \{Y_j \in C, D_z = j, D_{z'} \neq j\} - Q \{Y_j \in C, D_{z'} = j, D_z \neq j\} \\
&\leq Q \{Y_j \in B, D_z = j, D_{z'} \neq j\}~,
\end{align*}
from which the validity follows.

Conversely, we show that the same $Q^\ast$ in the proof of Theorem \ref{thm:sharp-Y} attains the right-hand side of \eqref{eq:subs} for every Borel set $B \subseteq \mathcal Y$ and all indices as stated, and hence \eqref{eq:subs} holds with equality. We will use the notation from the proof of Theorem \ref{thm:sharp-Y}, where we have already established $Q^\ast \in \mathbf Q^{\mathsf Y}$ and $P = Q^\ast T^{-1}$. In terms of its Radon-Nikodym derivatives, the supremum in \eqref{eq:subs} equals
\[ \int_B \big ( p_j(y | z) - p_j(y | z') \big )_+ d\nu(y)~, \]
attained by the Borel set $C = B \cap \{y : p_j(y | z) \geq p_j(y | z')\}$.

First suppose $z, z' \in \mathcal Z(j)$. Fix $y_j$, and let $i$ and $i'$ be the positions of $z$ and $z'$ in the ordering from that proof, so that $z_i(y_j,j) = z$ and $z_{i'}(y_j,j) = z'$. In block $j$ of the construction of $Q^\ast$, the condition $d_z = j$ holds exactly in steps $1, \dots, i$, whereas $d_{z'} \neq j$ holds exactly in steps $i' + 1, \dots, |\mathcal Z(j)|$. Therefore, the required pattern occurs in steps $i' + 1,\dots,i$ if $i > i'$, and in no step if $i \leq i'$. When $i > i'$, the successive differences assigned in these steps sum to $p_j(y_j | z) - p_j(y_j | z')$; when $i \leq i'$, then by definition of $z_i(y_j, j)$ and $z_{i'}(y_j, j)$, this difference is less than or equal to zero. Therefore,
\[ \sum_{(d_w : w \in \mathcal Z) \in \mathcal S_j: d_z = j, d_{z'} \neq j} q^\ast_{(d_w : w \in \mathcal Z)}(y_0, \dots, y_{J - 1}) = \prod_{j' \neq j} \lambda_{j'}(y_{j'}) \big ( p_j(y_j | z) - p_j(y_j | z') \big )_+~. \]
The equality holds when either $z$ or $z'$ is $\mathsf{base}$ as well, since the construction puts $\mathsf{base}$ last in every ordering. No vector in other blocks or on the diagonal has $d_w = j$ for $w \neq j$. Integrating over $y_j \in B$ and $y_{j'} \in \mathcal Y$ for $j' \neq j$, and recalling $\int_{\mathcal Y} \lambda_{j'}(y) d\nu(y) = 1$, we have
\[ Q^\ast \{Y_j \in B, D_z = j, D_{z'} \neq j\} = \int_B \big ( p_j(y | z) - p_j(y | z') \big )_+ d\nu(y)~. \]

It remains to consider $z = j$ or $z' = j$. In this case, the support of $\mathcal Z$ implies $J_0 \leq j \leq J - 1$. If $z' = j$, then it follows from Assumption \ref{as:default} that $D_z = j$ implies $D_j = j$, so both sides of \eqref{eq:subs} are zero. If $z = j$, then it follows from Assumption \ref{as:default} that $D_{z'} = j$ implies $D_j = j$. As a result,
\begin{align*}
& Q^\ast \{Y_j \in B, D_j = j, D_{z'} \neq j\} \\
&= P \{Y \in B, D = j | Z = j\} - P \{Y \in B, D = j | Z = z'\} \\
&= \int_B \big ( p_j(y | j) - p_j(y | z') \big )_+ d\nu(y)~,
\end{align*}
where the last equality follows from \eqref{eq:encourage-Y}. \qed

\subsection{Proof of Corollary \ref{cor:tv}}
Fix $z, z' \in \mathcal Z$ such that $z \neq z'$. For each $0 \leq j \leq J - 1$, Theorem \ref{thm:subs} with $B = \mathcal Y$ implies
\begin{align*}
& \min_{Q \in \mathbf Q^{\mathsf Y}: P = Q T^{-1}} Q \{D_z = j,D_{z'} \neq j\} \\
& \hspace{3em} = \sup_{C \subseteq \mathcal Y} \big ( P \{Y \in C,D = j | Z = z\} - P \{Y \in C,D = j | Z = z'\} \big )~,    
\end{align*}
where the supremum is over Borel sets $C \subseteq \mathcal Y$. The events $\{D_z = j,D_{z'} \neq j\}$, $0 \leq j \leq J - 1$ are disjoint and their union is $\{D_z \neq D_{z'}\}$. Moreover, Theorem \ref{thm:subs} shows that the same distribution $Q^\ast$ constructed in the proof of Theorem \ref{thm:sharp-Y} attains all of the preceding minima simultaneously. Therefore,
\begin{align}
\nonumber & \min_{Q \in \mathbf Q^{\mathsf Y}: P = Q T^{-1}} Q \{D_z \neq D_{z'}\}  \\
\label{eq:tv-proof} & = \sum_{0 \leq j \leq J - 1} \sup_{C \subseteq \mathcal Y} \big ( P \{Y \in C,D = j | Z = z\} - P \{Y \in C,D = j | Z = z'\} \big )~.
\end{align}
It remains to show that the right-hand side of \eqref{eq:tv-proof} is the total variation distance in \eqref{eq:tv-switching}. Let
\[ \mu_z(A) = P\{(Y, D) \in A | Z = z\} \]
for Borel sets $A \subseteq \mathcal Y \times \{0, \dots, J - 1\}$. For any such $A$ and $0 \leq j \leq J - 1 $, define
\[ A_j = \{y \in \mathcal Y: (y, j) \in A\}~. \]
Then,
\[ \mu_z(A) - \mu_{z'}(A) = \sum_{0 \leq j \leq J - 1} \big ( P \{Y \in A_j,D = j | Z = z\} - P \{Y \in A_j,D = j | Z = z'\} \big )~. \]
Because the sets $A_j$ can be chosen independently across $j$,
\begin{align*}
& \sup_A (\mu_z(A) - \mu_{z'}(A)) \\
& \hspace{3em} = \sum_{0 \leq j \leq J - 1} \sup_{C \subseteq \mathcal Y} \big ( P \{Y \in C, D = j | Z = z\} - P \{Y \in C, D = j | Z = z'\} \big )~.   
\end{align*}

Finally, because $\mu_z$ and $\mu_{z'}$ are both probability measures,
\[ \sup_A (\mu_z(A) - \mu_{z'}(A) ) = \sup_A |\mu_z(A) - \mu_{z'}(A)| = \|\mu_z - \mu_{z'}\|_{\mathrm{TV}}~. \]
Combining this equality with \eqref{eq:tv-proof}, we get \eqref{eq:tv-switching}. The simultaneous-attainment statement follows directly from the corresponding statement in Theorem \ref{thm:subs}. \qed

\subsection{Proof of Corollary \ref{cor:sources}}
Fix $Q \in \mathbf Q^{\mathsf Y}$ such that $P = Q T^{-1}$ and $k,\ell \in \{J_0, \dots, J - 1\}$ such that $k \neq \ell$. By Assumption \ref{as:default}, $D_\ell = k$ implies $D_k = k$. Therefore,
\[ Q \{D_k = k, D_\ell \neq k\} = Q \{D_k = k\} - Q \{D_\ell = k\}~, \]
and \eqref{eq:entry} follows from Assumption \ref{as:exog-Y}.

Next, fix $0 \leq j \leq J - 1$ such that $j \notin \{k, \ell\}$. Because $Q \in \mathbf Q^{\mathsf Y}$ satisfies Assumption \ref{as:default} and $j \notin \{k, \ell\}$, if $D_\ell = j$, then $j$ is the default, so $D_k \in \{k, j\}$. Therefore, with probability one under $Q$,
\[ \{D_\ell = j, D_k = k\} = \{D_\ell = j, D_k \neq j\}~. \]
Intersecting this equality with $\{Y_j \in B\}$ and setting $j \notin \{k, \ell\}$, $z = \ell$ and $z' = k$ in Theorem \ref{thm:subs}, we obtain \eqref{eq:source}.

Finally, noting that $\{D_\ell \neq k, D_k = k\} = \bigcup_{0 \leq j \leq J - 1: j \neq k} \{D_\ell = j, D_k = k\}$, we have from \eqref{eq:entry} that
\begin{align*}
& Q \{D_\ell = \ell, D_k = k\} \\
& = P \{D = k | Z = k\} - P \{D = k | Z = \ell\} - \sum_{0 \leq j \leq J - 1: j \notin \{k, \ell\}} Q \{D_\ell = j, D_k = k\}~,
\end{align*}
so \eqref{eq:pairwise-complier} follows from applying \eqref{eq:source} with $B = \mathcal Y$ to each term in the sum, noting the minima are attained simultaneously by $Q^\ast$ in the proof of Theorem \ref{thm:sharp-Y}. \qed

\subsection{Proof of Theorem \ref{thm:onesided}}
As shown in the paragraph after Theorem \ref{thm:onesided}, the conditions of Corollary \ref{cor:0better-Y} hold because of the stated inequalities and the support restrictions $P \{D = 0 | Z = \mathsf{base}\} = 1$ and $P \{D = j | Z = z\} = 0$ for $1 \leq z \leq J - 1$ and $j \notin \{z, 0\}$. Here, we provide a direct proof for completeness. Necessity follows from \eqref{eq:onesided} and Assumption \ref{as:exog-Y}. To prove sufficiency, we apply the construction in the proof of Theorem \ref{thm:sharp-Y} with $\mathcal Z = \{\mathsf{base}, 1, \dots, J - 1\}$. The support restrictions allow us to take $p_j(y | z) = 0$ for $1 \leq j \leq J - 1$ and $z \in \mathcal Z(j)$. Therefore, for $1 \leq j \leq J - 1$, if $P \{D = j | Z = j\} > 0$, the density in \eqref{eq:lambda} becomes
\[ \lambda_j(y) = \frac{p_j(y | j)}{P \{D = 0 | Z = \mathsf{base}\} - P \{D = 0 | Z = j\}} = \frac{p_j(y | j)}{P \{D = j | Z = j\}}~, \]
and if $P \{D = j | Z = j\} = 0$, then we define $\lambda_j(y) \equiv 1$ (or any measurable and nonnegative function such that $\int_{\mathcal Y} \lambda_j(y) d \nu(y) = 1$). The constructed distribution satisfies $Q^\ast \{D_{\mathsf{base}} = 0\} = P \{D = 0 | Z = \mathsf{base}\} = 1$. Because $Q^\ast$ is also constructed to satisfy both \eqref{eq:default} and \eqref{eq:default-base} in Assumption \ref{as:default}, it satisfies \eqref{eq:onesided}. Therefore, $Q^\ast \in \mathbf Q_{\mathsf{1s}}^{\mathsf{Y}}$. \qed

\subsection{Proof of Lemma \ref{lem:cells}}
We begin with the following lemma that shows that Theorem \ref{thm:sharp-Y} continues to hold if we replace the Borel $\sigma$-algebra with any algebra of sets that generates it.

\begin{lemma} \label{lem:algebra}
Let $\mathcal A$ be an algebra of subsets of $\mathcal Y$ and suppose $\mathcal A$ generates the Borel $\sigma$-algebra on $\mathcal Y$. Then, Theorem \ref{thm:sharp-Y}, Corollary \ref{cor:0better-Y}, and Theorem \ref{thm:onesided} continue to hold if every occurrence of ``Borel set'' is replaced by ``set in $\mathcal A$.''
\end{lemma}

\begin{proof}
We start with Theorem \ref{thm:sharp-Y}(a). As in the proof of Theorem \ref{thm:sharp-Y}, define a probability measure $\nu$ on $\mathcal B(\mathcal Y)$ by
\[ \nu(B) = \frac{1}{|\mathcal Z|} \sum_{z \in \mathcal Z} P \{Y \in B | Z = z\}~. \]
For Borel sets $A, B \subseteq \mathcal Y$, $0 \leq j \leq J - 1$, and $z \in \mathcal Z$, it follows from the triangle inequality that
\begin{equation} \label{eq:measure-continuity}
|P \{Y \in A, D = j | Z = z\} - P \{Y \in B, D = j | Z = z\}| \leq |\mathcal Z| \nu(A \triangle B)~,
\end{equation}
where $A \triangle B$ denotes the symmetric difference between $A$ and $B$.

We first show that for every Borel set $B \subseteq \mathcal Y$, there is a sequence of sets $A_n \in \mathcal A$ such that $\nu(A_n \triangle B) \to 0$. To that end, define
\[ \mathcal M = \{B \in \mathcal B(\mathcal Y): \exists \{A_n\}_{n \geq 1} \subseteq \mathcal A \text{ such that } \nu(A_n \triangle B) \to 0\}~. \]
Trivially, $\mathcal A \subseteq \mathcal M$. Next, we show $\mathcal M$ is a monotone class. Suppose $B_n \in \mathcal M$ and $B_n \uparrow B$. Then, for each $B_n$ there exists a set $A_n$ such that $\nu(A_n \triangle B_n) < 1 / n$. Because $A_n \triangle B \subseteq (A_n \triangle B_n) \cup (B_n \triangle B)$, it follows from Bonferroni's inequality that $\nu(A_n \triangle B) \to 0$. Therefore, $B \in \mathcal M$. Similarly, if $B_n \in \mathcal M$ and $B_n \downarrow B$, then $B \in \mathcal M$. So indeed $\mathcal M$ is a monotone class. Because $\mathcal A$ is an algebra, it follows from the monotone class theorem \citep[Theorem 4.4.2 in][]{dudley2002real} that $\mathcal M = \mathcal B(\mathcal Y)$.

Next, fix a collection of Borel sets $\{B_z(j): z \in \mathcal Z(j)\}$, $0 \leq j \leq J - 1$, such that $\{B_z(j): z \in \mathcal Z(j)\}$ forms a partition of $\mathcal Y$ for each $j$ as in Theorem \ref{thm:sharp-Y}(a). For each $j$, enumerate $\mathcal Z(j) = \{z_1, \dots, z_K\}$ for $K = |\mathcal Z(j)|$, and choose $A_{i, n} \in \mathcal A$ with $\nu(A_{i, n} \triangle B_{z_i}(j)) \to 0$ for $1 \leq i \leq K - 1$, and recall $B_{z_K}(j) = \mathcal Y \setminus \bigcup_{1 \leq s \leq K - 1} B_{z_s}(j)$. For each $n \geq 1$, define
\begin{align*}
B_{z_i}^{(n)}(j) & = A_{i, n} \setminus \bigcup_{1 \leq s \leq i - 1} A_{s, n} \text{ for } 1 \leq i \leq K - 1 \\
B_{z_K}^{(n)}(j) & = \mathcal Y \setminus \bigcup_{1 \leq s \leq |\mathcal Z(j)| - 1} A_{s, n}~.
\end{align*}
These sets form a partition of $\mathcal Y$, including when $K = 1$, and each of them belongs to $\mathcal A$. We now show that for all $z \in \mathcal Z(j)$,
\begin{equation} \label{eq:symdiff}
B_z^{(n)}(j) \triangle B_z(j) \subseteq \bigcup_{1 \leq i \leq K - 1} (A_{i, n} \triangle B_{z_i}(j))   
\end{equation}
Indeed, if $y \in \mathcal Y \setminus \bigcup_{1 \leq i \leq K - 1} (A_{i,n} \triangle B_{z_i}(j))$, then $I\{y \in A_{i, n}\} = I\{y \in B_{z_i}(j)\}$ for all $1 \leq i \leq K - 1$. Because $B_{z_i}(j), 1 \leq i \leq K$ form a partition of $\mathcal Y$, if $y \in B_{z_i}(j)$ for some $1 \leq i \leq K - 1$, then $y \in A_{i, n}$ and $y \notin A_{s, n}$ for all $1 \leq s \leq K - 1$, $s \neq i$, so $y \in B_{z_i}^{(n)}(j)$. If $y \in B_{z_K}(j)$, then $y \notin A_{i, n}$ for all $1 \leq i \leq K - 1$, so $y \in B_{z_K}^{(n)}(j)$. In any case, $y \notin B_z^{(n)}(j) \triangle B_z(j)$ for all $z \in \mathcal Z(j)$. Therefore, \eqref{eq:symdiff} holds. As a result,
\begin{equation} \label{eq:symdiff-nu}
\nu \big ( B_z^{(n)}(j) \triangle B_z(j) \big ) \leq \sum_{1 \leq i \leq K - 1} \nu ( A_{i,n} \triangle B_{z_i}(j)) \to 0~.
\end{equation}
By assumption, the inequalities in \eqref{eq:ineq-Y} hold with $\{B_z^{(n)}(j): z \in \mathcal Z(j)\}$, $0 \leq j \leq J - 1$. Together with \eqref{eq:measure-continuity} and \eqref{eq:symdiff-nu}, it implies the inequalities in \eqref{eq:ineq-Y} also hold for $\{B_z(j): z \in \mathcal Z(j)\}$, $0 \leq j \leq J - 1$.

Next, suppose the inequalities in \eqref{eq:encourage-Y} in Theorem \ref{thm:sharp-Y}(b) hold for all $A \in \mathcal A$. Again for each Borel set $B \subseteq \mathcal Y$, there exists a sequence $A_n \in \mathcal A$ such that $\nu(A_n \triangle B) \to 0$. Applying \eqref{eq:measure-continuity} to both sides of \eqref{eq:encourage-Y}, we get \eqref{eq:encourage-Y} for $B$. Similarly, Corollary \ref{cor:0better-Y} and Theorem \ref{thm:onesided} continue to hold if the conditions therein hold for all sets in $\mathcal A$ rather than all Borel sets.
\end{proof}

\begin{proof}[\sc Proof of Lemma \ref{lem:cells}]
Let $\mathcal A_m$ be the collection of all unions of sets in $\mathcal I_m$, and define $\mathcal A = \bigcup_{m \geq 1} \mathcal A_m$.
Because $\mathcal I_{m + 1}$ is a refinement of $\mathcal I_m$ for every $m \geq 1$, $\mathcal A_m \subseteq \mathcal A_{m + 1}$, and it can be verified that $\mathcal A$ is an algebra of Borel subsets of $\mathcal Y$. Fix a relatively open set $G \subseteq \mathcal Y$ and fix $y \in G$. Then, there exists $\epsilon > 0$ such that $\{u \in \mathcal Y: \|u - y\| < \epsilon\} \subseteq G$, where $\|\cdot\|$ denotes the Euclidean norm. Because $\mathrm{diam}(I_m(y)) \to 0$ for all $y \in \mathcal Y$ and $I_m(y) \subseteq \mathcal Y$, it follows that if $m$ is large enough so that $\mathrm{diam}(I_m(y)) < \epsilon$, then $I_m(y) \subseteq G$. Therefore,
\[ G = \bigcup_{m \geq 1} \bigcup_{I \in \mathcal I_m:\ I \subseteq G} I~, \]
a countable union of sets in $\mathcal A$. So $\mathcal A$ generates $\mathcal B(\mathcal Y)$, and Lemma \ref{lem:algebra} applies.

It remains to establish that the inequalities in \eqref{eq:ineq-Y} for sets in $\mathcal A$ are equivalent to the inequalities in \eqref{eq:ineq-cells}. Note that each set in $\mathcal A_m$ for each $m \geq 1$ is a union of sets in $\mathcal I_m$. Suppose first that the conditions in Theorem \ref{thm:sharp-Y}(a) hold for sets in $\mathcal A$. Fix $m \geq 1$ and a collection of $z(j, I) \in \mathcal Z(j)$ for $0 \leq j \leq J - 1$ and $I \in \mathcal I_m$. For each $0 \leq j \leq J - 1$ and $z \in \mathcal Z(j)$, define
\[ B_z(j) = \bigcup_{I \in \mathcal I_m: z(j, I) = z} I~. \]
Then, $\{B_z(j): z \in \mathcal Z(j)\}$ forms a partition of $\mathcal Y$ for each $j$, where each $B_z(j) \in \mathcal A$ because $\mathcal A$ is an algebra. Therefore, Theorem \ref{thm:sharp-Y}(a) for sets in $\mathcal A$ implies \eqref{eq:ineq-cells}. Conversely, suppose \eqref{eq:ineq-cells} holds for every $m \geq 1$. Fix a collection of sets in $\mathcal A$, $\{B_z(j): z \in \mathcal Z(j)\}$, $0 \leq j \leq J - 1$, such that $\{B_z(j): z \in \mathcal Z(j)\}$ forms a partition of $\mathcal Y$ for each $j$. Because $(\mathcal A_m: m \geq 1)$ is nested, there exists $M$ such that every $B_z(j) \in \mathcal A_M$ for all $0 \leq j \leq J - 1$ and $z \in \mathcal Z(j)$. Each $B_z(j)$ is therefore a union of cells in $\mathcal I_M$, and hence \eqref{eq:ineq-cells} with $m = M$ and $z(j, I) = z$ for $I \subseteq B_z(j)$ implies
\[ \sum_{0 \leq j \leq J - 1} \sum_{z \in \mathcal Z(j)} P \{Y \in B_z(j), D = j | Z = z\} \leq 1~. \]
Therefore, Theorem \ref{thm:sharp-Y}(a) holds with sets in $\mathcal A$, which establishes the desired equivalence.

Finally, consider any of the inequalities in \eqref{eq:encourage-Y}, \eqref{eq:0better-Y}, or \eqref{eq:onesided-testable}. If it holds for every $m \geq 1$ and $I \in \mathcal I_m$, then by finite additivity it holds for every set in $\mathcal A_m$ for every $m \geq 1$, and hence for every set in $\mathcal A$. Lemma \ref{lem:algebra} then implies that it holds for every Borel set. The converse is immediate.
\end{proof}

\section{Additional Details for the Examples}

\cite{kirkeboen2016field} study the effects of fields of study on earnings. In their setting, $D \in \{0, 1, 2\}$ represents three fields of study, ordered by their (soft) admission cutoffs from the lowest to the highest. The instrument $Z \in \{\mathsf{base}, 1, 2\}$ is called ``expected offer'' in their paper. Roughly speaking, $Z = 1$ when the student crosses the (soft) admission cutoff for field 1, $Z = 2$ when the student crosses the (soft) admission cutoff for field 2, and $Z = \mathsf{base}$ otherwise. The authors assume that $Z$ is exogenous in the sense that $Q$ satisfies Assumption \ref{as:exog}. In addition, they impose the following monotonicity conditions:
\begin{align}
    \label{eq:monotonicity1} Q \{D_1 = 1 | D_{\mathsf{base}} = 1\} & = 1~, \\
    \label{eq:monotonicity2} Q \{D_2 = 2 | D_{\mathsf{base}} = 2\} & = 1~.
\end{align}
The conditions in \eqref{eq:monotonicity1}--\eqref{eq:monotonicity2} require that crossing the cutoff for field 1 or 2 weakly encourages them towards that field. They further impose the following ``irrelevance'' conditions:
\begin{align}
    \label{eq:irrelevance1} Q \{I \{D_1 = 2\} & = I \{D_{\mathsf{base}} = 2\} | D_{\mathsf{base}} \neq 1, D_1 \neq 1\} = 1~, \\
    \label{eq:irrelevance2} Q \{I \{D_2 = 1\} & = I \{D_{\mathsf{base}} = 1\} | D_{\mathsf{base}} \neq 2, D_2 \neq 2\} = 1~.
\end{align}
The condition in \eqref{eq:irrelevance1} states that if crossing the cutoff for field 1 does not cause the student to switch to field 1, then it does not cause them to switch to or away from field 2. A similar interpretation applies to \eqref{eq:irrelevance2}. We now show the restrictions in \eqref{eq:monotonicity1}--\eqref{eq:irrelevance2} imply Assumption \ref{as:default} with $J = 3$ and $\mathcal Z = \{\mathsf{base}, 1, 2\}$, as the other direction is straightforward. Suppose by contradiction that $D_1 \notin \{D_{\mathsf{base}}, 1\}$. In that case, if $D_1 = 2$, then \eqref{eq:monotonicity1} implies that $D_{\mathsf{base}} \neq 1$; \eqref{eq:irrelevance1} therefore implies that $D_{\mathsf{base}} = 2$, so $D_1 = D_{\mathsf{base}}$, a contradiction to $D_1 \notin \{D_{\mathsf{base}}, 1\}$. If instead $D_1 = 0$, then \eqref{eq:monotonicity1} again implies that $D_{\mathsf{base}} \neq 1$; because $D_1 \notin \{D_{\mathsf{base}}, 1\}$, we know $D_{\mathsf{base}} = 2$, which by \eqref{eq:irrelevance1} implies $D_1 = 2$, a contradiction to $D_1 = 0$. Therefore, \eqref{eq:default} holds for $j = 1$ with $\boldsymbol{j}^{\ast} = D_{\mathsf{base}}$, and hence \eqref{eq:default-base} holds as well. Similar arguments show it is also satisfied for $j = 2$.

The support of $(D_{\mathsf{base}}, D_1, D_2)$ is summarized in Table \ref{table:klm}. The restrictions may be violated if, for instance, the change from $Z = \mathsf{base}$ to $Z = 1$ not only increases the appeal of field 1 but also decreases the appeal of field 0.

{
\spacingset{\TABLESPACING}
\begin{table}[ht!]
    \centering
    \begin{tabular}{ccccccccc}
    \toprule
    $D_{\mathsf{base}}$ & 0 & 0 & 0 & 1 & 1 & 2 & 2 & 0 \\
    $D_1$ & 0 & 1 & 0 & 1 & 1 & 2 & 1 & 1 \\
    $D_2$ & 0 & 0 & 2 & 1 & 2 & 2 & 2 & 2 \\
    \bottomrule
    \end{tabular}
    \caption{The support of $(D_{\mathsf{base}}, D_1, D_2)$ in \cite{kirkeboen2016field}.}
    \label{table:klm}
\end{table}
}

\cite{kirkeboen2016field} derive causal interpretations of the IV estimand under the assumptions above plus the restriction that $D_{\mathsf{base}} = 0$, which they call the ``next best'' condition. The full set of assumptions is then equivalent to one-sided noncompliance, meaning that $D_{\mathsf{base}} = 0$ and $D_j \in \{j, 0\}$ for $j \in \{1, 2\}$. Under all these assumptions together with Assumption \ref{as:exog-Y}, they show that an IV regression identifies the average treatment effects for the ``compliers'' of each instrument value relative to $Z = \mathsf{base}$. Our results in Section \ref{sec:onesided} apply to the case when the ``next best'' condition is additionally imposed.

Next, we provide details for the assumptions made in \cite{behaghel2013robustness}. They impose the following two requirements which they refer to as ``extended monotonicity'':
\begin{align}
\label{eq:em-1} Q \{I \{D_2 = 1\} = I \{D_0 = 1\} & \leq I \{D_1 = 1\}\} = 1 \\
\label{eq:em-2} Q \{I \{D_1 = 2\} = I \{D_0 = 2\} & \leq I \{D_2 = 2\}\} = 1~.
\end{align}
We argue that \eqref{eq:em-1}--\eqref{eq:em-2} imply Assumption \ref{as:default} with $J = 3$ and $\mathcal Z = \{\mathsf{base}, 1, 2\}$, where $\mathsf{base}$ is relabeled to $0$, and are strictly stronger. Indeed, if $D_0 = 1$, then $D_1 = 1$ and $D_2 = 1$ by \eqref{eq:em-1}. If $D_0 = 2$, then $D_2 = 2$ and $D_1 = 2$ by \eqref{eq:em-2}. If $D_0 = 0$, then $D_2 \neq 1$ by \eqref{eq:em-1}, so $D_2 \in \{0, 2\} = \{D_0, 2\}$, and similarly $D_1 \in \{0, 1\} = \{D_0, 1\}$ by \eqref{eq:em-2}. In all cases, $D_j \in \{D_0, j\}$ for $0 \leq j \leq 2$. Clearly, \eqref{eq:em-1}--\eqref{eq:em-2} are stronger than the base-state specification because, under the same relabeling, $(D_0, D_1, D_2) = (1, 1, 2)$ satisfies Assumption \ref{as:default} but violates \eqref{eq:em-1}.

\section{Equivalence with a Random Utility Model} \label{sec:rum}
Although Assumption \ref{as:default} is intuitive, we now provide additional motivation for it by showing that the model that imposes Assumptions \ref{as:exog}--\ref{as:default} is \emph{equivalent} to a fairly general random utility model, and such a model will shed more light on how the default choice naturally arises. The same model was shown by \cite{kline2016evaluating} to imply the restriction in \eqref{eq:kw}. We will in turn show that they are equivalent. We emphasize, however, that the results in this appendix only serve as motivation for Assumption \ref{as:default} and our results in the main text do not rely on the random utility model.

Fix $0 \leq J_0 \leq J - 1$. Recall that $\mathcal Z$ is either $\{J_0, \dots, J - 1\}$ or $\{\mathsf{base}\} \cup \{J_0, \dots, J - 1\}$, with $|\mathcal Z| \geq 2$. Let $\Omega$ denote the underlying probability space. To define the random utility model, for $0 \leq j \leq J - 1$ and $z \in \mathcal Z$, let $U_j(z, \cdot): \Omega \to \mathbf R$ be a random variable that denotes the random utility of choice $j$ when the instrument equals $z$. We assume that $(U_j(z, \cdot): 0 \leq j \leq J - 1, z \in \mathcal Z) \indep Z$. The random utility model maintains that the potential treatments $D_z$ for $z \in \mathcal Z$ are given by
\begin{equation} \label{eq:rum-potential}
D_z(\omega) \in \argmax_{0 \leq j \leq J - 1}~ U_j(z, \omega)~.
\end{equation}
Ties in \eqref{eq:rum-potential} are resolved by choosing the smallest $j$ that solves \eqref{eq:rum-potential}. Therefore, each $D_z$ is a measurable function of the random utilities, and the selection rule does not depend on the realized instrument $Z$.

To capture the idea that $Z = j$ only encourages towards $D = j$, we make the following restriction on the random utilities. For $0 \leq j \leq J - 1$, let $\ubar U_j(\cdot)$ be a random variable that denotes the baseline utility of choice $j$. We assume that $U_j(z, \omega) = \ubar U_j(\omega)$ for all $0 \leq j \leq J - 1$ and $z \in \mathcal Z(j)$, and $U_j(j, \omega) \geq \ubar U_j(\omega)$ for $J_0 \leq j \leq J - 1$.
The restriction captures the idea that the utilities of the first $J_0$ choices are not affected by the instrument, while for $J_0 \leq j \leq J - 1$, the utility is (weakly) increased when $z = j$ but otherwise stays the same. When $\mathsf{base} \in \mathcal Z$, we assume $U_j(\mathsf{base}, \omega) = \ubar U_j(\omega)$ for $0 \leq j \leq J - 1$. To rule out degenerate situations, we will further assume that the distribution of $(\ubar U_j(\omega): 0 \leq j \leq J - 1)$ is absolutely continuous with respect to the Lebesgue measure on $\mathbf R^J$.

Let $\mathbf Q_{\mathrm{rum}}$ denote our random utility model; that is, for every $Q \in \mathbf Q_{\mathrm{rum}}$, 
(a) $D_z$ is determined by \eqref{eq:rum-potential}, with the tie-breaking rule specified above; 
(b) the distribution of $(\ubar U_j(\cdot): 0 \leq j \leq J - 1)$ is absolutely continuous with respect to the Lebesgue measure on $\mathbf R^J$;
(c) $U_j(z, \omega) = \ubar U_j(\omega)$ for $0 \leq j \leq J - 1$ and $z \in \mathcal Z(j)$, and $U_j(j, \omega) \geq \ubar U_j(\omega)$ for $J_0 \leq j \leq J - 1$; if $\mathsf{base} \in \mathcal Z$, then $U_j(\mathsf{base}, \omega) = \ubar U_j(\omega)$ for $0 \leq j \leq J - 1$;
and (d) $(U_j(z, \cdot): 0 \leq j \leq J - 1, z \in \mathcal Z) \indep Z$. 
The following result shows that $\mathbf Q_{\mathrm{rum}} = \mathbf Q$, so that each $Q$ that satisfies the encouragement design restriction on potential treatments can be generated by the random utility model under consideration. It extends \cite{vytlacil2002independence} to our setting, although a crucial distinction is that the instrument is discrete in our setting, and therefore the proof is elementary.

\begin{lemma} \label{lem:default}
$\mathbf Q_{\mathrm{rum}} = \mathbf Q$.
\end{lemma}

\begin{proof}[\textup{\textsc{Proof of Lemma \ref{lem:default}}}]
\underline{$\mathbf Q_{\mathrm{rum}} \subseteq \mathbf Q$}. To show that $\mathbf Q_{\mathrm{rum}} \subseteq \mathbf Q$, we need to show that every $Q$ in the random utility model satisfies Assumptions \ref{as:exog} and \ref{as:default}. Assumption \ref{as:exog} follows because $(D_z : z \in \mathcal Z)$ is a measurable function of the random utilities and therefore independent of $Z$. Let $\Omega$ denote the underlying probability space. For each $\omega \in \Omega$, let
\begin{equation} \label{eq:default-rum}
\boldsymbol{j}^{\ast}(\omega) \in \argmax_{0 \leq j \leq J - 1}~ \ubar U_j(\omega)~.    
\end{equation}
In case of ties, we choose the smallest $j$ that solves \eqref{eq:default-rum}. $\boldsymbol{j}^{\ast}$ defined this way is a random variable because it is determined by comparisons of a finite number of random utilities. Because the distribution of $(\ubar U_j(\cdot): 0 \leq j \leq J - 1)$ is absolutely continuous with respect to the Lebesgue measure on $\mathbf R^J$, $\boldsymbol{j}^{\ast}$ is unique with probability one. Here, $\boldsymbol{j}^{\ast}(\omega)$ is the default choice when the instrument does not exist, and the value of $\boldsymbol{j}^{\ast}(\omega)$ could differ across $\omega \in \Omega$. Furthermore, because ties happen with probability zero, we know that with probability one,
\begin{equation} \label{eq:default-utility}
\ubar U_{\boldsymbol{j}^{\ast}(\omega)}(\omega) > \max_{k \neq \boldsymbol{j}^{\ast}(\omega)} \ubar U_k(\omega)~.
\end{equation}
Next, we show by contradiction that with probability one, for $J_0 \leq j \leq J - 1$, $D_j(\omega) \in \{j, \boldsymbol{j}^{\ast}(\omega)\}$. Indeed, suppose $D_j(\omega) = k \notin \{j, \boldsymbol{j}^{\ast}(\omega)\}$ for a set of $\omega$ with strictly positive probability. Then,
\begin{equation} \label{eq:contra}
U_k(j, \omega) \geq U_{\boldsymbol{j}^{\ast}(\omega)}(j, \omega)~.    
\end{equation}
If $j \neq \boldsymbol{j}^{\ast}(\omega)$, then \eqref{eq:contra} implies that
\[ \ubar U_k(\omega) = U_k(j, \omega) \geq U_{\boldsymbol{j}^{\ast}(\omega)}(j, \omega) = \ubar U_{\boldsymbol{j}^{\ast}(\omega)}(\omega) \]
with strictly positive probability, a contradiction to \eqref{eq:default-utility}. If $j = \boldsymbol{j}^{\ast}(\omega)$, then \eqref{eq:contra} implies
\[ \ubar U_k(\omega) = U_k(j, \omega) \geq U_{\boldsymbol{j}^{\ast}(\omega)}(\boldsymbol{j}^{\ast}(\omega), \omega) \geq \ubar U_{\boldsymbol{j}^{\ast}(\omega)}(\omega)~, \]
another contradiction to \eqref{eq:default-utility}. Therefore, for $J_0 \leq j \leq J - 1$, we have shown with probability one that $D_j(\omega) \in \{j, \boldsymbol{j}^{\ast}(\omega)\}$. If $\mathsf{base} \in \mathcal Z$, then $U_j(\mathsf{base}, \omega) = \ubar U_j(\omega)$ for $0 \leq j \leq J - 1$, so
\[ D_{\mathsf{base}}(\omega) \in \argmax_{0 \leq j \leq J - 1} \ubar U_j(\omega)~, \]
and therefore \eqref{eq:default-rum} and the absolute continuity of the distribution of $(\ubar U_j(\cdot): 0 \leq j \leq J - 1)$ imply that $D_{\mathsf{base}} = \boldsymbol{j}^{\ast}$ with probability one.

\underline{$\mathbf Q \subseteq \mathbf Q_{\mathrm{rum}}$}. To show $\mathbf Q \subseteq \mathbf Q_{\mathrm{rum}}$, we need to show that every distribution $Q$ that satisfies Assumptions \ref{as:exog}--\ref{as:default} can be generated from a random utility model in $\mathbf Q_{\mathrm{rum}}$. To do so, we will show that $Q$ can be generated by an \emph{additive} random utility model $\mathbf Q_{\mathrm{rum}}' \subseteq \mathbf Q_{\mathrm{rum}}$ which is seemingly much more restrictive. Doing so will imply $\mathbf Q \subseteq \mathbf Q_{\mathrm{rum}}' \subseteq \mathbf Q_{\mathrm{rum}}$, and because we already know $\mathbf Q_{\mathrm{rum}} \subseteq \mathbf Q$, these inclusions become equalities; that is, $\mathbf Q = \mathbf Q_{\mathrm{rum}}' = \mathbf Q_{\mathrm{rum}}$, and the proof is concluded.

To define the additive random utility model $\mathbf Q_{\mathrm{rum}}'$, let $\delta_0, \dots, \delta_{J - 1}: \mathcal{Z} \to \mathbf R$ be deterministic functions and let $(\epsilon_0, \dots, \epsilon_{J - 1})$ be a random vector of unobservables, the distribution of which is absolutely continuous with respect to the Lebesgue measure on $\mathbf R^J$. Further suppose $(\epsilon_0, \dots, \epsilon_{J - 1})$ and $Z$ are independent. The additive random utility model maintains that the potential treatments $D_z$ for $z \in \mathcal{Z}$ are given by
\begin{equation} \label{eq:enc-potential}
D_z \in \argmax_{0 \leq j \leq J - 1}~ (\delta_j(z) + \epsilon_j)~.
\end{equation}
Because the distribution of $(\epsilon_0, \dots, \epsilon_{J - 1})$ is absolutely continuous with respect to the Lebesgue measure on $\mathbf R^J$, ties happen with probability zero. Therefore, with probability one, $D_z$ is unique and
\[ I \{D_z = j\} = I \{\delta_j(z) + \epsilon_j > \delta_k(z) + \epsilon_k \text{ for all } k \neq j\}~. \]
We further assume that
\[ \delta_j(z) = \alpha_j + \beta_j I \{z = j\} \]
for some $\beta_0, \dots, \beta_{J - 1} \geq 0$. In other words, each instrument value other than $\mathsf{base}$ encourages a unique choice. Because $\alpha_j$ can be absorbed into $\epsilon_j$ without loss of generality, we henceforth assume $\alpha_j = 0$, and accordingly
\begin{equation} \label{eq:specific}
\delta_j(z) = \beta_j I \{z = j\}~.
\end{equation}

When $J_0 > 0$, to reflect the idea that the first $J_0$ choices are not affected by the instrument, we simply set $\beta_j = 0 \text{ for all } j < J_0$. If $\mathsf{base} \in \mathcal Z$, then $\delta_j(\mathsf{base}) = 0$ for all $0 \leq j \leq J - 1$, so the base state does not shift any utility. Fix $0 \leq J_0 \leq J - 1$. Let $\mathbf Q_{\mathrm{rum}}'$ denote the set of all distributions of $((D_z: z \in \mathcal Z), Z)$ which are consistent with our additive random utility model. That is, under each $Q \in \mathbf Q_{\mathrm{rum}}'$, 
(a) $D_z$ is determined by \eqref{eq:enc-potential}; 
(b) the distribution of $(\epsilon_0, \dots, \epsilon_{J - 1})$ is absolutely continuous with respect to the Lebesgue measure on $\mathbf R^J$; 
(c) $\delta_0, \dots, \delta_{J - 1}$ are given by \eqref{eq:specific}, where $\beta_j = 0$ for $j < J_0$ and $\beta_j \geq 0$ for $J_0 \leq j \leq J - 1$; 
and (d) $(\epsilon_0, \dots, \epsilon_{J - 1}) \indep Z$. It is clear that $\mathbf Q_{\mathrm{rum}}' \subseteq \mathbf Q_{\mathrm{rum}}$ with $\epsilon_j(\omega) = \ubar U_j(\omega)$.

Set $\beta_j = 0$ if $j < J_0$ and $\beta_j = 1$ if $J_0 \leq j \leq J - 1$. Assumption \ref{as:default} can equivalently be expressed as requiring the probabilities of some vectors of potential treatments are zero. In particular, define
\[ \mathcal S = \{(d_z : z \in \mathcal Z) \in \{0, \dots, J - 1\}^{\mathcal Z}: d_z \neq z,\ d_k \neq k \implies d_z = d_k \text{ for all } z, k \in \mathcal Z\}~. \]
Assumption \ref{as:default} can then be expressed as $Q \{(D_z: z \in \mathcal Z) = (d_z: z \in \mathcal Z)\} = 0$ if $(d_z: z \in \mathcal Z) \notin \mathcal S$. This is true even if $\mathsf{base} \in \mathcal Z$. For $(d_z: z \in \mathcal Z) \in \mathcal S$, we define $\boldsymbol{j}^{\ast}(d_z: z \in \mathcal Z)$ as the associated default choice, which is the value of $d_z$ as long as $d_z \neq z$. If $\mathsf{base} \notin \mathcal Z$ and $d_z = z$ for all $z \in \mathcal Z$, then we define $\boldsymbol{j}^{\ast}(d_z: z \in \mathcal Z) = 0$ as a convention. 

For $(d_z: z \in \mathcal Z) \in \mathcal S$, define
\[ \Lambda(d_z: z \in \mathcal Z) = \{J_0 \leq j \leq J - 1: d_j = j \neq \boldsymbol{j}^{\ast}(d_z: z \in \mathcal Z)\}~. \]
$\Lambda$ consists of the encouragements that make a person switch away from the default. We will sometimes suppress the dependence of $\boldsymbol{j}^{\ast}$ and $\Lambda$ on $(d_z: z \in \mathcal Z)$ for convenience. Set $M = 2$. For $(d_z: z \in \mathcal Z) \in \mathcal S$, define the corresponding region for $(\epsilon_0, \epsilon_1, \dots, \epsilon_{J - 1})$ as
\begin{align*}
R(d_z: z \in \mathcal Z) = \{(\epsilon_0, \dots, \epsilon_{J - 1}): ~&\epsilon_{\boldsymbol{j}^{\ast}} > \epsilon_j \text{ for } j \neq \boldsymbol{j}^{\ast}, \\
&\beta_j + \epsilon_j > \epsilon_k \text{ for } j \in \Lambda 
\text{ and } k \neq j, \\
&\beta_j + \epsilon_j < \epsilon_{\boldsymbol{j}^{\ast}} \text{ for } j \notin \Lambda \text{ and } j \neq \boldsymbol{j}^{\ast}, \\
&|\epsilon_j| \leq M \text{ for } 0 \leq j \leq J - 1 \} \subseteq \mathbf R^J~.
\end{align*}
For $j < J_0$, the third inequality is redundant because $\beta_j = 0$.
For every $(d_z: z \in \mathcal Z) \in \mathcal S$, we have $|R(d_z: z \in \mathcal Z)| > 0$, where $|R(d_z: z \in \mathcal Z)|$ denotes the Lebesgue measure of $R(d_z: z \in \mathcal Z)$. To see this, take $\epsilon_{\boldsymbol{j}^{\ast}} = 0$, $\epsilon_j = -1/2$ for $j \in \Lambda$, and $\epsilon_j = -3/2$ for $j \neq \boldsymbol{j}^{\ast}$ and $j \notin \Lambda$. All the inequalities hold strictly at this point, and $|\epsilon_j| < 2$ for every $j$. By continuity, they remain strict in a sufficiently small open box around the point, so the region has positive volume; its volume is finite because it is bounded. Next, every point in $R(d_z: z \in \mathcal Z)$ leads to $(d_z: z \in \mathcal Z)$. The first inequality in its definition makes $\boldsymbol{j}^{\ast}$ the unique maximizer of $(\epsilon_0, \dots, \epsilon_{J - 1})$. If $\mathsf{base} \in \mathcal Z$, then $\delta_j(\mathsf{base}) + \epsilon_j = \epsilon_j$ for $0 \leq j \leq J - 1$, so $D_{\mathsf{base}} = \boldsymbol{j}^{\ast} = d_{\mathsf{base}}$. For $J_0 \leq j \leq J - 1$, if $j = \boldsymbol{j}^{\ast}$, then $j$ remains uniquely optimal when $Z = j$ because $\beta_j \geq 0$; if $j \neq \boldsymbol{j}^{\ast}$ and $j \in \Lambda$, the second inequality makes $j$ uniquely optimal when $Z = j$; if instead $j \neq \boldsymbol{j}^{\ast}$ and $j \notin \Lambda$, then the third inequality guarantees $\delta_j(j) + \epsilon_j < \delta_{\boldsymbol{j}^{\ast}}(j) + \epsilon_{\boldsymbol{j}^{\ast}}$, so $D_j = \boldsymbol{j}^{\ast}$. In particular, $R(d_z: z \in \mathcal Z)$ for distinct $(d_z: z \in \mathcal Z)$ are disjoint.

Let $(\epsilon_0, \dots, \epsilon_{J - 1})$ be uniformly distributed in each $R(d_z: z \in \mathcal Z)$, with density
\[ \frac{Q\{(D_z: z \in \mathcal Z) = (d_z: z \in \mathcal Z) \}}{|R(d_z: z \in \mathcal Z)|} \]
and let its density be zero outside of the union of these regions. The density integrates to one. Let $Q^\ast$ denote the distribution of $((D_z: z \in \mathcal Z), Z)$ generated by $(\epsilon_0, \dots, \epsilon_{J - 1})$ and an independently generated $Z$ from the marginal distribution of $Z$ under $Q$. Then, Assumption \ref{as:exog} implies $Q^\ast = Q$. Therefore, $Q \in \mathbf Q_{\mathrm{rum}}'$, which proves the reverse inclusion and completes the proof.
\end{proof}

\section{Details for Simulations} \label{sec:sims-appdx}

\subsection{Finite Outcomes} \label{sec:sims-finite-null}
For each specification, we verify the values of $\theta$ that belong to the null hypothesis in Section \ref{sec:sims} by checking the full set of sharp inequalities. First consider the binary outcome $Y = I \{0 < V \leq 2\}$ in Section \ref{sec:sims-finite}. Recall $a = 2 \Phi(1/2) - 1$, $b = 2 \Phi(2) - 1$, and $c_\ast = \frac{a}{3(b - a)}$. Define $q_\theta = a + \theta(b - a)$. Then, $P \{Y = 1 | D = 0, Z = j\} = q_\theta$ for $1 \leq j \leq J - 1$, while $P \{Y = 1 | D = d, Z = z\} = a$ for all other cells $(d, z)$ with positive probability. Since $\mathcal Y = \{0, 1\}$, each pairwise inequality for all Borel sets is equivalent to its two singleton inequalities for $\{0\}$ and $\{1\}$.

We first consider the specifications without and with a base state, for which $z_0 = 0$ and $z_0 = \mathsf{base}$. In both cases, for $1 \leq j \leq J - 1$, a necessary inequality is
\[ P \{Y = 1, D = 0 | Z = j\} \leq P \{Y = 1, D = 0 | Z = z_0\}~. \]
Without a base state, this is an instance of \eqref{eq:encourage-Y}; with a base state, it is an instance of \eqref{eq:0better-Y}. The difference between its two sides is
\begin{equation} \label{eq:sims-binary-difference}
P \{Y = 1, D = 0 | Z = j\} - P \{Y = 1, D = 0 | Z = z_0\} = \frac{3 \theta(b - a) - a}{4J}~.
\end{equation}
Therefore, each of these inequalities holds if and only if $\theta \leq c_\ast$. For the other outcome value, we have
\[ P \{Y = 0, D = 0 | Z = j\} - P \{Y = 0, D = 0 | Z = z_0\} = -\frac{1 - a + 3 \theta(b - a)}{4 J} \leq 0~. \]
For $d \geq 1$, the conditional distribution of $Y \mid D = d, Z = z$ does not depend on $z$, so the remaining pairwise inequalities follow from the corresponding inequalities on the choice probabilities. In particular, with a base state, \eqref{eq:encourage-Y} and \eqref{eq:0better-Y} hold if and only if $\theta \leq c_\ast$.

Without a base state, where $J = 3$, it remains to check the partition inequalities \eqref{eq:ineq-Y}. Because $Y$ is binary, it is straightforward to show these inequalities hold if and only if
\[ \sum_{y \in \{0, 1\}} \sum_{d = 0}^2 \max_{z \in \mathcal Z(d)} P \{Y = y, D = d | Z = z\} \leq 1~. \]
For $d = 0$, both eligible instrument values have the same conditional outcome distribution, so the two maxima are $q_\theta / 4$ for $y = 1$ and $(1 - q_\theta) / 4$ for $y = 0$. For $d = 1$ and $d = 2$, both maxima are attained at $z = 0$, and their sum over $y$ is $1/3$. Therefore,
\[
\sum_{y \in \{0, 1\}} \sum_{d = 0}^2
\max_{z \in \mathcal Z(d)} P \{Y = y, D = d | Z = z\}
= \frac{1}{4} + \frac{2}{3} = \frac{11}{12} < 1~.
\]
Therefore, the partition inequalities hold for every $\theta \in [0, 1]$. Together with the preceding comparisons, this calculation shows that \eqref{eq:ineq-Y} and \eqref{eq:encourage-Y} hold if and only if $\theta \leq c_\ast$ in the specification without a base state.

Finally, consider one-sided noncompliance with $J \in \{3, 4\}$. The support restrictions in Theorem \ref{thm:onesided} hold by construction. The probabilities of $D = 0$ are now 1 under $Z = \mathsf{base}$ and $3/4$ under every other instrument value. Therefore, for $1 \leq j \leq J - 1$, the two displayed probability differences above are multiplied by $J$. These comparisons exhaust the inequalities in \eqref{eq:onesided-testable} for the binary outcome. It follows that \eqref{eq:onesided-testable} holds for every Borel set if and only if $\theta \leq c_\ast$.

\subsection{Continuous Outcomes} \label{sec:sims-continuous-null}
We now consider $Y = V$ in the specification with $\mathsf{base} \in \mathcal Z$ and $J = 4$, as in Section \ref{sec:sims-continuous}. Let $f$ and $g$ denote the densities of $N(1, 4)$ and $N(1, 1/4)$ and let $f_\theta = (1 - \theta) f + \theta g$. For $1 \leq j \leq J - 1$, the conditional outcome distribution $Y \mid D = 0, Z = j$ has density $f_\theta$; all other cells with positive probability have density $f$. Because $P \{Y \in B, D = d | Z = z\} = P \{Y \in B | D = d, Z = z\} P \{D = d | Z = z\}$, for comparisons in \eqref{eq:encourage-Y} and \eqref{eq:0better-Y} that have the same conditional outcome distribution on both sides, it suffices to compare the choice probabilities. The inequalities in \eqref{eq:encourage-Y} hold because, for $d \in \{1, 2, 3\}$, its choice probability is the largest under $Z = d$. The inequalities in \eqref{eq:0better-Y} also hold except possibly for $D = 0$ under $Z = j$ versus $Z = \mathsf{base}$, for $1 \leq j \leq J - 1$.

The remaining conditions are that, for every Borel set $B \subseteq \mathbf R$ and $1 \leq j \leq J - 1$,
\[ P \{Y \in B, D = 0 | Z = j\} \leq P \{Y \in B, D = 0 | Z = \mathsf{base}\}~. \]
For $j \in \{1, 2, 3\}$, the measures on the two sides have densities $3 f_\theta / (4J)$ and $f / J$. Therefore, these conditions are equivalent to
\begin{equation} \label{eq:sims-density-domination}
\frac{3}{4} f_\theta(v) \leq f(v) \text{ for Lebesgue-almost every } v \in \mathbf R~.
\end{equation}
Because
\[ \frac{g(v)}{f(v)} = 4 \exp \bigg ( -\frac{15}{8} (v - 1)^2 \bigg )~, \]
we have
\[ \sup_{v \in \mathbf R} \frac{3 f_\theta(v)}{4 f(v)} = \frac{3}{4} (1 + 3 \theta)~. \]
It follows that \eqref{eq:sims-density-domination} holds if $\theta \leq 1/9$. If $\theta > 1/9$, however, it fails throughout an open interval containing 1, by continuity of the densities. Taking this interval as $B$ violates \eqref{eq:0better-Y}. Hence, the full set of inequalities \eqref{eq:encourage-Y} and \eqref{eq:0better-Y} holds if and only if $\theta \leq 1/9$.

Next, consider the two fixed partitions in Section \ref{sec:sims-continuous}. Write $F(I) = \int_I f(v) dv$ and $G(I) = \int_I g(v) dv$. For either partition, all the cell inequalities hold if and only if
\[ \frac{3}{4} \big ( (1 - \theta) F(I) + \theta G(I) \big ) \leq F(I) \text{ for every cell } I~. \]
For $\mathcal I_1 = \{(-\infty, 1], (1, \infty)\}$, we have $F(I) = G(I) = 1/2$ in both cells, so these inequalities hold for every $\theta \in [0, 1]$. For $\mathcal I_2 = \{(-\infty, 0], (0, 1], (1, 2], (2, \infty)\}$, each central interval has $F(I) = a/2$ and $G(I) = b/2$. Its inequality is therefore equivalent to
\[ \frac{3}{8} \big ( a + \theta(b - a) \big ) \leq \frac {a}{2}~, \]
which holds if and only if $\theta \leq c_\ast$. Each tail interval has $F(I) = (1 - a)/2$ and $G(I) = (1 - b)/2 < F(I)$, so its inequality holds for every $\theta \in [0, 1]$. Hence, the four-bin inequalities hold if and only if $\theta \leq c_\ast$. In particular, when $1/9 < \theta \leq c_\ast$, the four-bin distribution is consistent with the model even though the full continuous-outcome distribution is not.

\subsection{Additional Results}

Table \ref{table:sims-onesided} reports simulation results for one-sided noncompliance in Section \ref{sec:sims-finite}.

{
\spacingset{\TABLESPACING}
\begin{table}[ht!]
    \centering\normalsize
    \begin{tabular}{rcccccc@{\hspace{8pt}}cccccc}
    \toprule
    \multicolumn{13}{c}{One-sided noncompliance} \\
    \addlinespace[4pt]
    \multicolumn{13}{c}{(a) $J = 3$} \\
    \addlinespace[4pt]
    & \multicolumn{6}{c}{\textbf{Closed Form}} & \multicolumn{6}{c}{\textbf{Latent}} \\
    \cmidrule(lr){2-7} \cmidrule(lr){8-13}
    & & \multicolumn{2}{c}{$H_0$} & \multicolumn{3}{c}{$H_1$} & & \multicolumn{2}{c}{$H_0$} & \multicolumn{3}{c}{$H_1$} \\
    \cmidrule(lr){3-4} \cmidrule(lr){5-7} \cmidrule(lr){9-10} \cmidrule(lr){11-13}
    $n$ &  & $\theta = 0$ & $c_\ast$ & $0.30$ & $0.40$ & $0.50$ &  & $0$ & $c_\ast$ & $0.30$ & $0.40$ & $0.50$ \\
    \cmidrule(lr){1-13}
    500 &  & 0.0 & 6.2 & 18.5 & 49.9 & 80.8 &  & 0.0 & 5.6 & 16.3 & 46.4 & 78.4 \\
    1000 &  & 0.0 & 9.1 & 30.5 & 74.6 & 95.6 &  & 0.0 & 7.3 & 27.7 & 72.4 & 95.4 \\
    2000 &  & 0.0 & 8.2 & 43.4 & 93.8 & 100.0 &  & 0.0 & 6.3 & 40.4 & 93.1 & 100.0 \\
    \addlinespace[12pt]
    \multicolumn{13}{c}{(b) $J = 4$} \\
    \addlinespace[4pt]
    & \multicolumn{6}{c}{\textbf{Closed Form}} & \multicolumn{6}{c}{\textbf{Latent}} \\
    \cmidrule(lr){2-7} \cmidrule(lr){8-13}
    & & \multicolumn{2}{c}{$H_0$} & \multicolumn{3}{c}{$H_1$} & & \multicolumn{2}{c}{$H_0$} & \multicolumn{3}{c}{$H_1$} \\
    \cmidrule(lr){3-4} \cmidrule(lr){5-7} \cmidrule(lr){9-10} \cmidrule(lr){11-13}
    $n$ &  & $\theta = 0$ & $c_\ast$ & $0.30$ & $0.40$ & $0.50$ &  & $0$ & $c_\ast$ & $0.30$ & $0.40$ & $0.50$ \\
    \cmidrule(lr){1-13}
    500 &  & 0.0 & 5.8 & 15.9 & 43.0 & 71.5 &  & 0.0 & 5.2 & 14.4 & 40.2 & 68.9 \\
    1000 &  & 0.0 & 7.5 & 27.4 & 65.7 & 94.2 &  & 0.0 & 6.4 & 23.3 & 61.5 & 93.2 \\
    2000 &  & 0.0 & 9.8 & 38.9 & 89.3 & 99.8 &  & 0.0 & 8.0 & 33.8 & 87.6 & 99.7 \\
    \bottomrule
    \end{tabular}
    \caption{rejection rates (\%) for binary outcomes under one-sided noncompliance.}
    \label{table:sims-onesided}
\end{table}
}

To illustrate the computational advantage of the closed-form inequalities for a finer outcome partition, we extend the continuous-outcome specification in Section \ref{sec:sims-continuous} to eight bins. We use the refinement
\[ \mathcal I_3 = \{(-\infty, -1.5], (-1.5, 0], (0, 0.5], (0.5, 1], (1, 1.5], (1.5, 2], (2, 3.5], (3.5, \infty)\}~. \]
Table \ref{table:sims-eight} reports rejection probabilities at the 5\% level, using 100 replications and 500 bootstrap draws. The eight-bin inequalities remain strictly slack at $\theta = 1/9$. Closed Form and Latent use the same observations and bootstrap draws within each replication.

{
\spacingset{\TABLESPACING}
\begin{table}[ht!]
    \centering\normalsize
    \setlength{\tabcolsep}{3.5pt}
    \begin{tabular}{rlcccccc@{\hspace{8pt}}cccccc}
    \toprule
    \multicolumn{14}{c}{$\mathsf{base} \in \mathcal Z$, $J = 4$} \\
    \addlinespace[4pt]
     &  & \multicolumn{6}{c}{\textbf{Closed Form}} & \multicolumn{6}{c}{\textbf{Latent}} \\
    \cmidrule(lr){3-8} \cmidrule(lr){9-14}
     &  &  & \multicolumn{2}{c}{$H_0$} & \multicolumn{3}{c}{$H_1$} &   & \multicolumn{2}{c}{$H_0$} & \multicolumn{3}{c}{$H_1$} \\
    \cmidrule(lr){4-5} \cmidrule(lr){6-8} \cmidrule(lr){10-11} \cmidrule(lr){12-14}
    $n$ &  &  & $\theta = 0$ & $1/9$ & $0.50$ & $0.75$ & $1$ &  & $0$ & $1/9$ & $0.50$ & $0.75$ & $1$ \\
    \cmidrule(lr){1-14}
2000 & Eight bins &  & 0.0 & 0.0 & 19.0 & 72.0 & 98.0 &  & 0.0 & 0.0 & 8.0 & 46.0 & 87.0 \\
    \bottomrule
    \end{tabular}
    \caption{Rejection probabilities (\%) for continuous outcomes with eight bins, $\mathsf{base} \in \mathcal Z$, $J = 4$. Hypotheses concern the full distribution of $Y$.}
    \label{table:sims-eight}
\end{table}
}

Table \ref{table:sims-computation} reports matrix dimensions and mean runtime per test in seconds. For Closed Form, $A$ is the identity matrix in the slack formulation. Latent dimensions omit structural zeros and include normalization and target rows ($ + 2$), before redundant equations are removed.
For binary outcomes, four bins, and eight bins, we average the runtime over the values of $\theta$. The conditional probabilities without an outcome and with two bins do not change with $\theta$, so we compute each such test only once per replication and reuse its result for all $\theta$.

{
\spacingset{\TABLESPACING}
\begin{table}[ht!]
    \centering\normalsize
    \setlength{\tabcolsep}{2.5pt}
    \begin{tabular}{lcccc@{\hspace{8pt}}cccc}
    \toprule
    & \multicolumn{4}{c}{\textbf{Closed Form}} & \multicolumn{4}{c}{\textbf{Latent}} \\
    \cmidrule(lr){2-5} \cmidrule(lr){6-9}
    & & \multicolumn{3}{c}{Runtime (s)} & & \multicolumn{3}{c}{Runtime (s)} \\
    \cmidrule(lr){3-5} \cmidrule(lr){7-9}
    & $\dim(A)$ & $n = 500$ & $1000$ & $2000$ & $\dim(A)$ & $n = 500$ & $1000$ & $2000$ \\
    \addlinespace[6pt]
    \multicolumn{9}{l}{Finite outcome: $\mathsf{base} \notin \mathcal Z$, $J = 3$} \\
    \addlinespace[2pt]
     & $76\times76$ & 5.73 & 5.57 & 5.47 & $(18 + 2)\times80$ & 2.60 & 2.58 & 2.58 \\
    \addlinespace[6pt]
    \multicolumn{9}{l}{Finite outcome: $\mathsf{base} \in \mathcal Z$, $J = 4$} \\
    \addlinespace[2pt]
     & $36\times36$ & 4.64 & 4.43 & 4.32 & $(32 + 2)\times320$ & 3.31 & 3.25 & 3.23 \\
    \addlinespace[6pt]
    \multicolumn{9}{l}{One-sided noncompliance} \\
    \addlinespace[2pt]
    $J = 3$ & $4\times4$ & 4.08 & 3.98 & 3.91 & $(10 + 2)\times32$ & 2.45 & 2.42 & 2.39 \\
    $J = 4$ & $6\times6$ & 4.09 & 3.98 & 3.91 & $(14 + 2)\times128$ & 2.60 & 2.56 & 2.53 \\
    \addlinespace[6pt]
    \multicolumn{9}{l}{Continuous outcome: $\mathsf{base} \in \mathcal Z$, $J = 4$} \\
    \addlinespace[2pt]
    No outcome & $9\times9$ & 3.91 & 3.52 & 3.28 & $(16 + 2)\times20$ & 2.57 & 2.49 & 2.39 \\
    Two bins & $36\times36$ & 4.64 & 4.43 & 4.32 & $(32 + 2)\times320$ & 3.32 & 3.25 & 3.23 \\
    Four bins & $72\times72$ & 6.41 & 6.12 & 5.89 & $(64 + 2)\times5120$ & 11.84 & 11.95 & 12.06 \\
    Eight bins & $144\times144$ & - & - & 17.41 & $(128 + 2)\times81920$ & - & - & 349.78 \\
    \bottomrule
    \end{tabular}
    \caption{Matrix dimensions and mean runtime (seconds per test).}
    \label{table:sims-computation}
\end{table}
}

\clearpage
\bibliography{encouragement}
\end{document}